# Role of vibrations on decoherence in molecular spin qubits: The case of $[Cu(mnt)_2]^{2-}$

Luis Escalera-Moreno,[a] Nicolas Suaud,[b] Alejandro Gaita-Ariño*[a], and Eugenio Coronado[a]

**Abstract:** Herein we develop a simple first-principles methodology to determine the modulation that vibrations exert on spin energy levels, a key for the rational design of high–temperature molecular spin qubits and single–molecule magnets. This methodology is demonstrated by applying it to $[Cu(mnt)_2]^{2-}$ ($mnt^{2-}$ = 1,2-dicyanoethylene-1,2-dithiolate), a highly coherent complex, using DFT to calculate the normal vibrational modes and wave-function based theory calculations to estimate the spin energy level structure. By theoretically identifying the most relevant vibrational modes, we are able to offer general strategies to chemically design more resilient magnetic molecules, where the qubit energy is not coupled to local vibrations.

## Introduction

Quantum information technologies, which are in the spotlight of the latest European Flagship, are based on the detailed control of two-level quantum systems, known as quantum bits or qubits. Magnetic molecules are spin-unpaired quantum systems and thus they are an ideal platform to realize qubits. Moreover, they can be prepared and tailored by using well-known techniques of Molecular Magnetism, giving rise to the so-called molecular spin qubits.[1] Indeed, molecules hold great promise for the implementation of quantum algorithms,[2] and even for the scalability and organization of qubits.[3] Like other solid-state qubit candidates, they face the challenge of quantum decoherence: uncontrolled interactions with the environment that cause the loss of quantum information.

The three main sources of decoherence in molecular (electronic) spin qubits are: (a) the nuclear spin bath, (b) neighbouring electronic spins and (c) the phonon bath.[4] The first two are basically magnetic dipolar interactions and thus easier to determine from the point of view of physics.[5] Further, strategies to tune the nuclear spin bath are well developed and have recently allowed to reach a $T_2 \sim 0.7$ ms in strictly optimized conditions: a spin-free frozen solution of a $V^{IV}$ complex with spin-free ligands.[6]

On the contrary, decoherence arising from the coupling of the spin qubit states to phonons in complex molecular solids is, in general, the least well understood for the physics community since the details depend on the environment and the chemical structure of the compound under study. In particular, intramolecular vibrations can also couple to qubit states, and thus should also be tailored by synthetic chemistry. While they usually fall outside the low-frequency window that can interact with spin states in simple extended ionic lattices considered by physicists, these vibrations become relevant in energy in more complex molecular systems. This way, local vibrations are able to dominate that coupling by: (a) taking the place of acoustic phonons in Raman and Orbach[7(a)] or thermally activated[8] processes, which are also critically important for the relaxation of molecular nanomagnets, and (b) being the main contribution to the spin-lattice decoherence rate $1/T_1$ around liquid nitrogen temperatures.[9] Owing to the well-known bound $T_2 \leq 2T_1$, this means that the characteristic time $T_2$ will end up being controlled by the spin-lattice relaxation via molecular vibrations beyond those temperatures. Moreover, a general strategy to minimize spin bath decoherence, namely maximizing quantum tunnelling, dramatically increases vibrational decoherence,[4] as that strategy increases the density of resonant phonons with the energy separation of the spin doublet.

A chemical control and modelling of local vibrations is then necessary, whether the task is achieving a functional qubit or simply a molecular nanomagnet, if one wishes to operate at moderately high temperatures. Indeed, it is now accepted that engineering the crystal field splitting of highly axial anisotropic magnetic centers does not necessarily lead to an improved performance of single ion magnets.[7] Instead, there may appear other dominant relaxation processes that are not dependent on the magnitude of the energy barrier,[10] such as the spin-phonon coupling. In this sense, there are already authors claiming that both the role and tuning of spin-lattice/local vibrations interaction remain an open problem which impedes a clear and concise strategy for enhancing the properties either molecular spin qubits or single molecule magnets.[3(e),7]

From the experimental side, recent studies by R. Sessoli et al. have been shedding light on this issue. Vanadyl complexes acting as potential spin qubits has been explored at high temperatures,[11] evidencing the importance of molecular stiffness to enhance quantum coherence. In another previous study of a series of Cu(II) complexes with similar geometry, relaxations rates were higher for the flexible molecules than for the rigid ones. In that study, the fitting parameter $A_{loc}$ characterizing the contribution of local modes to the $T_1$-relaxation was taken as a quantitative measure of complex flexibility.[9]

Nonetheless, as those authors pointed out, an *ab-initio* framework able to encompass both rationalization and prediction of promising candidates is still missing. While phonon spectrum of ionic crystals is usually modeled according to the Debye model, its applicability is rather limited when dealing with complex molecular solids.[12] Furthermore, coefficients characterizing the different spin-phonon relaxation mechanisms are (a) usually challenging to calculate and thus extracted as fitting parameters, (b) sometimes assumed to be all equal in magnitude,[4] and (c) reduced in number when they are present in a large amount, as quantitative analysis would be hopeless otherwise.[7a]

In the present study, we will use $[Cu(mnt)_2]^{2-}$ (**1**, $mnt^{2-}$ = 1,2-dicyanoethylene-1,2-dithiolate, see Fig. 1)[13] as a model. This complex can be seen as a spin qubit in which the two states correspond to the ground spin doublet of $Cu^{2+}$, thus the qubit energy at a given field is proportional to the Landé factor $g$. In this system, a respectable $T_2 \sim 68$ μs was measured at low temperature (5 K) when the effect of the nuclear spin bath was minimized by deuteration. Above 100 K, $T_1$ and $T_2$ were both around 10 μs. This high coherence was attributed to the lattice rigidity and our goal is to take a step forward to quantify what such a rigidity means without the damage of the above limitations. The model here developed is able to incorporate any general discrete lattice or local vibration, once its harmonic frequency, reduced mass and vector displacement of atomic coordinates are known. For **1**, we will tackle only its intramolecular vibrational excitations. In particular, we will analyse both the population and the coupling of those excitations to the qubit transition energy with temperature. This analysis will allow us to identify the vibrational modes that may promote spin relaxation.

[a] L. Escalera-Moreno, A. Gaita-Ariño, E. Coronado
UIMM - ICMOL
Universidad de Valencia
José Beltrán 2, 46980, Paterna, Spain
luis.escalera@uv.es alejandro.gaita@uv.es
eugenio.coronado@uv.es
[b] N. Suaud
LCPQ - IRSAMC
Université Paul Sabatier
118 route de Narbonne, 31062, Toulouse, France
suaud@irsamc.ups-tlse.fr

In the next section, we will briefly present first the general features of the methodology, and then we will apply it to the case study, **1**. All additional technical details can be found as Supplementary Information (SI).

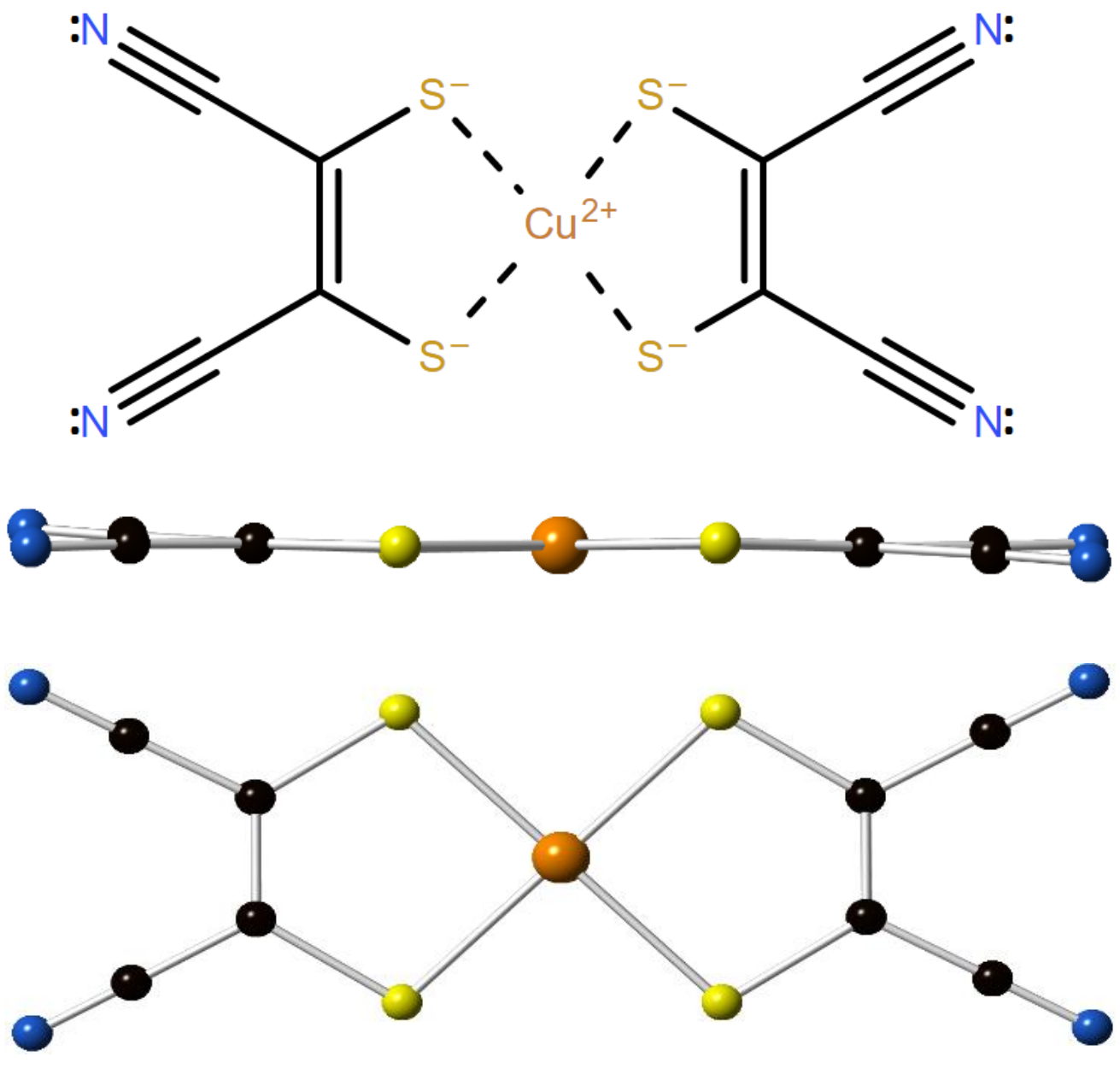

**Figure 1.** Top: Lewis structure showing also the coordination bonds (dashed lines). Middle and Bottom: Up and side views of the experimental geometry of **1** at 100 K. Orange: Copper, Yellow: Sulphur, Black: Carbon, Blue: Nitrogen. Note that the complex contains no H atoms.

## Results and Discussion

### Derivation of the model

The theoretical model starts by considering a property $B$ of our qubit whose alteration by vibrations affects its coherence. The influence of vibrations on that property, which will be modelled as harmonic, is accounted for by assuming that it is a function of $R$ generalized 1-dimensional vibrational coordinates $Q_k$: $B = B(Q_1,\ldots,Q_R)$. Then, a Taylor expansion up to second order around $Q_1 = \ldots = Q_R = 0$ is performed. At this point, a single-molecule $B$ expectation value is calculated under the harmonic approximation, allowing to quantify the property-vibration interaction with any and every vibrational mode. Thus, given a fixed set of harmonic vibrational quantum numbers $N = \{n_1,\ldots,n_R\}$, the harmonic vibrational wave function defined in Eq. (1) is expressed as the product of the $R$ harmonic wave functions of each vibrational mode:

$$\Psi^N(Q_1,\ldots,Q_R) = \prod_{k=1}^{R} \Psi_k^{n_k}(Q_k) \qquad (1)$$

If that Taylor expansion is used together with Eq. (1), such a single-molecule expectation value is obtained as a sum of $R$ independent contributions, one per each mode:

$$\langle B\rangle^N = \langle \Psi^N|B|\Psi^N\rangle \qquad (2)$$

$$\langle B\rangle^N \approx (B)_e + \frac{\hbar}{4\pi}\sum_{k=1}^{R}\left(\frac{\partial^2 B}{\partial Q_k^2}\right)_e \left(n_k + \frac{1}{2}\right)\frac{1}{\nu_k m_k} \qquad (3)$$

$(B)_e$ is the value of $B$ at $Q_1 = \ldots = Q_R = 0$, $\nu_k$ are the harmonic vibrational frequencies, $m_k$ are the reduced masses and the derivatives are evaluated at $Q_1 = \ldots = Q_R = 0$. Note that the model allows to deal with any rather general vibrational wave function, for example, to include anharmonicity, just by changing Eq. (1) and recalculating the expectation value, Eq. (2). In addition, one could extend that Taylor expansion up to third order, but because of parity arguments, since we are under the harmonic approximation, any odd-order term in Eq. (3) disappears and there are only even terms.

Finally, temperature is included by considering a Grand Canonical Ensemble, where the probability of each single-molecule expectation value, Eq. (3), characterized by the set $N = \{n_1,\ldots,n_R\}$, is given in terms of the Grand Partition Function $\mathcal{Z}$, see SI. Then, the following expression for the thermal dependence of the expectation value of B is obtained:

$$\overline{\langle B\rangle}(T) \approx \overline{\langle B\rangle}(T=0) + \sum_{k=1}^{R}\left[\frac{\hbar}{4\pi}\left(\frac{\partial^2 B}{\partial Q_k^2}\right)_e \frac{1}{m_k \nu_k}\langle n_k\rangle\right] \qquad (4)$$

where $\overline{\langle B\rangle}(T=0)$ and $\langle n_k\rangle$ are expressed as:

$$\overline{\langle B\rangle}(T=0) = (B)_e + \frac{\hbar}{8\pi}\sum_{k=1}^{R}\left[\left(\frac{\partial^2 B}{\partial Q_k^2}\right)_e \frac{1}{m_k \nu_k}\right] \qquad (5)$$

$$\langle n_k\rangle = \frac{1}{e^{\nu_k/k_B T} - 1} \qquad (6)$$

Eq. (5) defines the zero point contribution to $B$ and Eq. (6) is the boson number according to the Bose-Einstein statistics. This allows to estimate the modulation – and thus the effective coupling – of each vibrational mode in the property $B$ at any given temperature, as well as the total expectation value.

### Proposed methodology

The general procedure to follow when dealing with a specific system relies on three steps: (a) the relaxation of the geometry; (b) the determination of the vibrational modes under the harmonic approximation; and (c) the calculation of the second derivatives of the property $B$ of interest within an *ab-initio* approach.

As step (a), one relaxes the geometry of the set of atoms involving the relevant vibrations. If that set is a molecular complex, as experimental geometries in crystals are usually influenced by their surroundings, a relaxation in vacuum of this complex could give a geometry which is far from reality. Thus, it may be necessary to include also a portion of its closest surrounding environment in the relaxation. This was the case for **1**, where we kept the environment frozen and changed it by helium atoms; in order to model the steric effects onto the magnetic molecule. With this assumption, see SI, we got an acceptable relaxed geometry, which corresponds to $Q_1 = \ldots = Q_R = 0$ in terms of vibrational coordinates.

The step (b) consists in calculating the vibrational spectrum of the relaxed geometry. The relevant parameters to get from the ouput are the harmonic frequencies $\nu_k$, the reduced masses $m_k$ and the $3P$ dimensional displacement vectors $v_k$ of the vibrational modes, $P$ being the number of atoms.

Finally, in the step (c), the second derivative of $B$ for each vibrational mode $k$ is calculated. The key points to follow are: (1) generate a certain number of distorted geometries, around the optimized geometry; (2) calculate $B$ at each distorted geometry, by means of the suited *ab-initio* method; and (3) fit those calculated $B$ values to a proper polynomial, whose second derivative is analytically calculated at $Q_k = 0$, see Fig. 2. Each distorted geometry is given by the $3P$ dimensional vector, $v_{dist,k} = v_{eq} + Q_k \cdot v_k$, and is generated by giving $Q_k$ a proper value, see SI, being $v_{eq}$ the $3P$-dimensional vector of the atomic coordinates. The same set of distorted geometries would be generated employing $v_{dist,k} = v_{eq} - Q_k \cdot v_k$, where each geometry is obtained now with the same value of the distorsion coordinate $Q_k$ but with opposite sign. Note that the sign of the second derivative is preserved regardless the chosen criterion for the sign of the distortion coordinate, and thus it is well-defined.

For **1**, the set of atoms to optimize is the complex itself and the vibrations are its molecular modes, then $R = 3P\text{-}6$. Furthermore, we will assume that the parameter governing the qubit vibrational decoherence is its electron Landé $g_z$ factor, which directly determines the Zeeman splitting in practice between the two spin states of the ground $Cu^{2+}$ - S = 1/2, i.e., the qubit energy. Nevertheless, the methodology is analogous for any other molecular anisotropy parameter, such as the axial $D$ and rhombic $E$ zero field splittings, characteristic of molecular magnets; or the tunnelling gap $\Delta$ in anisotropic lanthanoid complexes, a crucial parameter for the design of lanthanoid-based spin qubits.[14]

## Some general considerations

When taking $B = g_z$ as a study parameter and considering the molecular modes as the relevant vibrations, Eq. (4) gives us $3P\text{-}6$ independent contributions $B_k$, Eq. (7), one per each vibrational mode:

$$B_k = \frac{\hbar}{4\pi}\left(\frac{\partial^2 g_z}{\partial Q_k^2}\right)_e \frac{1}{m_k \nu_k}\langle n_k \rangle \qquad (7)$$

These contributions should be as small as possible in order to make the qubit energy stable with temperature. In Eq. (7), the only temperature dependent factor is the boson number (Eq. (6)) which gives the thermal population of the mode. The remaining factor can be regarded as a constant that characterizes the coupling strength of said mode, Eq. (8). Thus, the total contribution is simply a balance between how populated the mode is and how strong that vibrational mode couples with the spin excitation. The best situation would be, of course, small couplings and low populations in all modes. As this is a rather exceptional case, the best balance to get will be having modes that hardly couple, although they could be significantly populated at the working temperature; or modes that, even if strongly coupled to the spin energy, are not highly populated. Synthetic efforts should be adressed to optimize such a balance in that direction, as recently explored involving vanadyl complexes.[11]

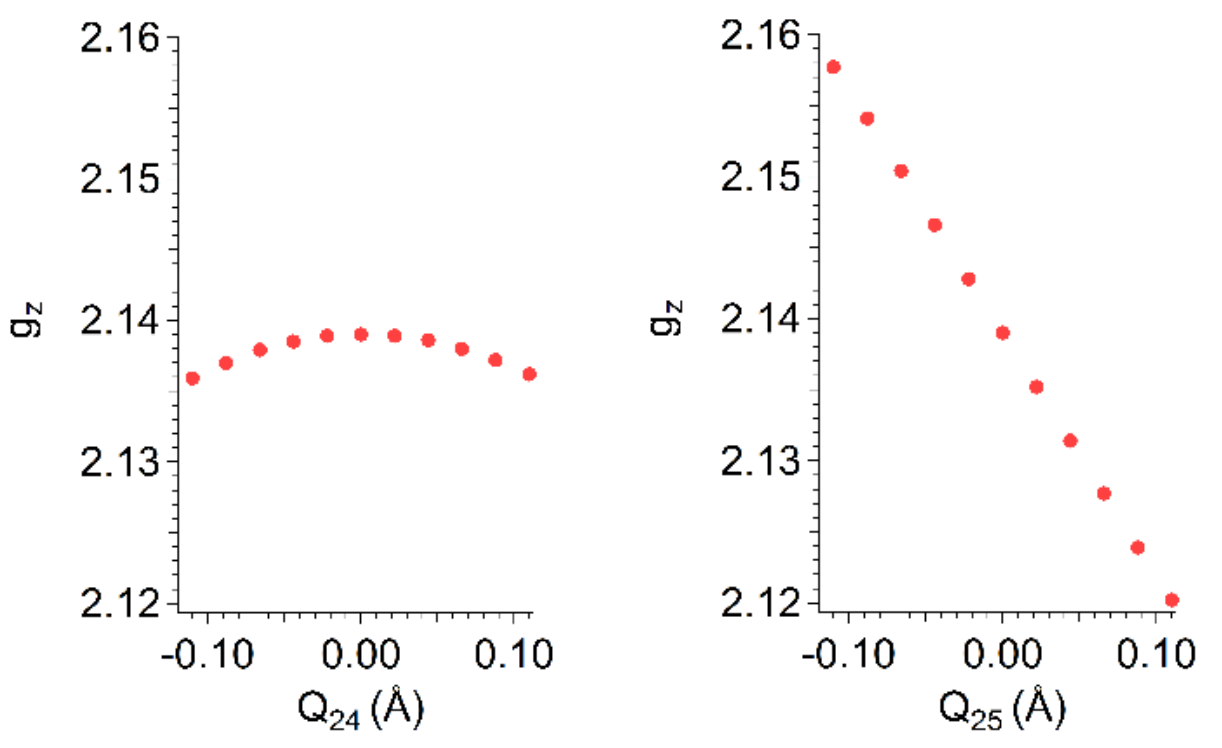

**Figure 2**. $g_z$ evolution along the distortion coordinate ranges of the modes 24 (left) and 25 (right) of **1** between -0.110 Å and +0.110 Å. The near linear dependence of $g_z$ with $Q_{25}$ gives a rather small second derivative. That vibrational movement corresponds to a molecular breathing, see Fig. 4.

$$C_k = \frac{\hbar}{4\pi}\left(\frac{\partial^2 g_z}{\partial Q_k^2}\right)_e \frac{1}{m_k \nu_k} \qquad (8)$$

Thus, given a reference molecular system and a working temperature, the strongly coupled and highly populated modes to suppress would be first identified, an achievable task through the present method. Then, rational chemical tuning would enter into play for vibrational optimization. The harmonic frequencies $\nu_k$ would be increased by making the complex as rigid as possible, for example, restricting the movement of those ligands involved in the relevant modes, as it happens in porphyrines or phtalocyanines; or trying to encapsulate the metallic ion inside a cage, resulting in a coordination environment as tight as possible. Systems like that could be fullerenes that host atoms[15,16], or “stapled” bis-phthalocyanines.[17] The reduced masses $m_k$ would be larger by removing light atoms, or replacing them by other heavier atoms but keeping low their spin-orbit coupling. In particular, one could replace hydrogen by deuterium or fluorine as done in [18]. For the second derivative to be as small as possible, two different strategies could be followed. First, to achieve a horizontal $g_z$ evolution with the distortion coordinate $Q_k$. This means having a highly isotropic electron spin (i.e., a very small spin-orbit coupling); this would be the case of organic radicals or, among metals, of $Gd^{3+}$ spin qubits.[19] Often, magnetic anisotropy is desirable since it facilitates qubit addressing;[20] in this case one needs to engineer vibrational modes whose action hardly affect the relevant molecular orbital energies contributing to $g_z$. This last situation can be encountered, for example, when vibrational modes and molecular orbitals are of different symmetry.[21] Second, a high molecular symmetry, which may permit to have modes with a linear $g_z$ evolution in the vibrational coordinate $Q_k$. This is the case of the mode 25 of **1**, a breathing vibration, in which $g_z$ evolves linearly along the distortion range, see Fig. 2. As a result, its second derivative takes a quite small value compared to other modes. Finally, in an ensemble of molecules, the boson number, Eq. (6), will always be smaller by cooling.

Therefore, identifying the relevant modes to suppress and knowing how they act on a given reference system, a vibrational optimization of series of compounds can then be proposed, allowing thus a chemical rational design of molecular spin qubits.[11(c)] As it will not be possible to independently optimize each relevant mode without affecting the others, one must look for the best balanced global situation.

## A case study: $[Cu(mnt)_2]^{2-}$

DFT (Density Functional Theory) was employed for the geometry optimization and for the vibrational spectrum calculation, using the package Gaussian09. The infrared spectrum of **1** was experimentally determined between 400 and 2500 $cm^{-1}$,[13] which allows checking the accuracy of our calculated spectrum in that energy range. Since we let only the complex **1** vibrate in its closest surrounding frozen, the displayed pics in Fig. 3 top and bottom (d) correspond to only-complex vibrations. This means that those peaks in bottom (c) corresponding to only-counterion vibrations do not appear in that calculated spectrum. For example, see the region around 750 $cm^{-1}$ in Fig. 3 bottom (c), where the 3 peaks are only-$PPh_4Br$ peaks. The marked peaks correspond to only-ligand peaks in Fig. 3 bottom (b), where the metal is not moved from its optimized position; or peaks involving both complex and $PPh_4Br$ peaks. Good estimates of those marked pics corresponding to infrared-active complex vibrations are satisfactorily reproduced, with errors of less than 5% (the maximum deviation appears in the highest energy vibration, where $\nu_{exp}$ = 2195 $cm^{-1}$, $\nu_{theo}$ = 2294 $cm^{-1}$, with the error being smaller for less energetic vibrations). This may serve as an indication of the expected accuracy of our final results. We attribute the non-perfect match of the calculated and experimental positions of the peaks mainly to the approximation to keep frozen the environment during the geometry relaxation of the complex. We are not aware of experimental IR spectrum below 400 cm$^{-1}$ available to compare with.

The calculated far-infrared IR spectrum of **1**, see Fig. 3 top, exhibits a first noticeable gap in between 305 cm$^{-1}$ and 390 cm$^{-1}$ (harmonic frequencies of the modes 25 and 26, respectively). These energies, in terms of temperature, are 439 K and 561 K. As experimental working temperatures range only up to room temperature, vibrations beyond such a gap will not be significantly populated, and thus we will only consider the first 25 vibrational modes in our calculations.

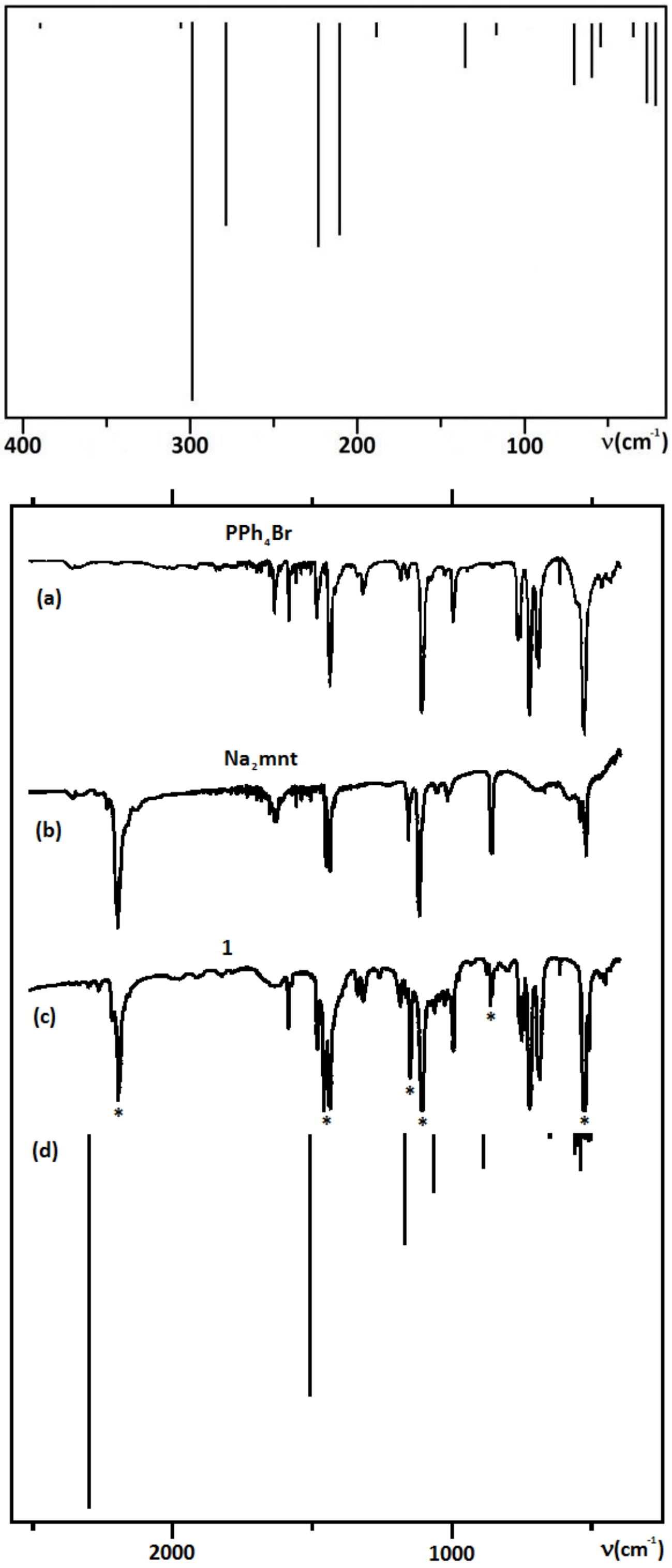

**Figure 3**. IR spectrum of **1**: Top: calculated in the far-infrared region at the optimized geometry; Bottom, experimental adapted from [13]: (a) only counter-ion, (b) only ligand mnt$^{2-}$, (c) complex + counter-ion, (d) calculated in the experimental frequency range at the optimized geometry. Vertical scale is in arbitrary units. The peaks marked in (c) are those involving only-ligand peaks in (b), or peaks involving both complex and counter-ion peaks.

A quick check of that subset of vibrations, as can be seen in some examples in Fig. 4, shows that they do not involve significant covalent bonds stretching (see SI for a selection of additional modes, including animated gifs). Instead, the most significant distortions consist in covalent bonds bending in the ligands, together with coordination bond stretching and bending. Often, the least energetic vibrations of the spectrum that can already be significantly populated from low temperature hardly alter bond angles or bond distances in the coordination sphere. As it will be seen below, this does not necessarily mean that they are not expected to modulate the spin energies. Let us consider for illustration the modes 4 and 12 (Fig 4). Mode 4, which is a low frequency mode and thus expected to populate at low *T*, is a twisting vibration that alters little the dihedral angle between the two ligands, which in this mode behave almost as rigid planes. Mode 12, which has a much higher frequency, can also be seen as a kind of twisting vibration within the ligands, which alters the orientation of the $CuS_4$ moiety which maintains its square planar structure (see SI for animation).

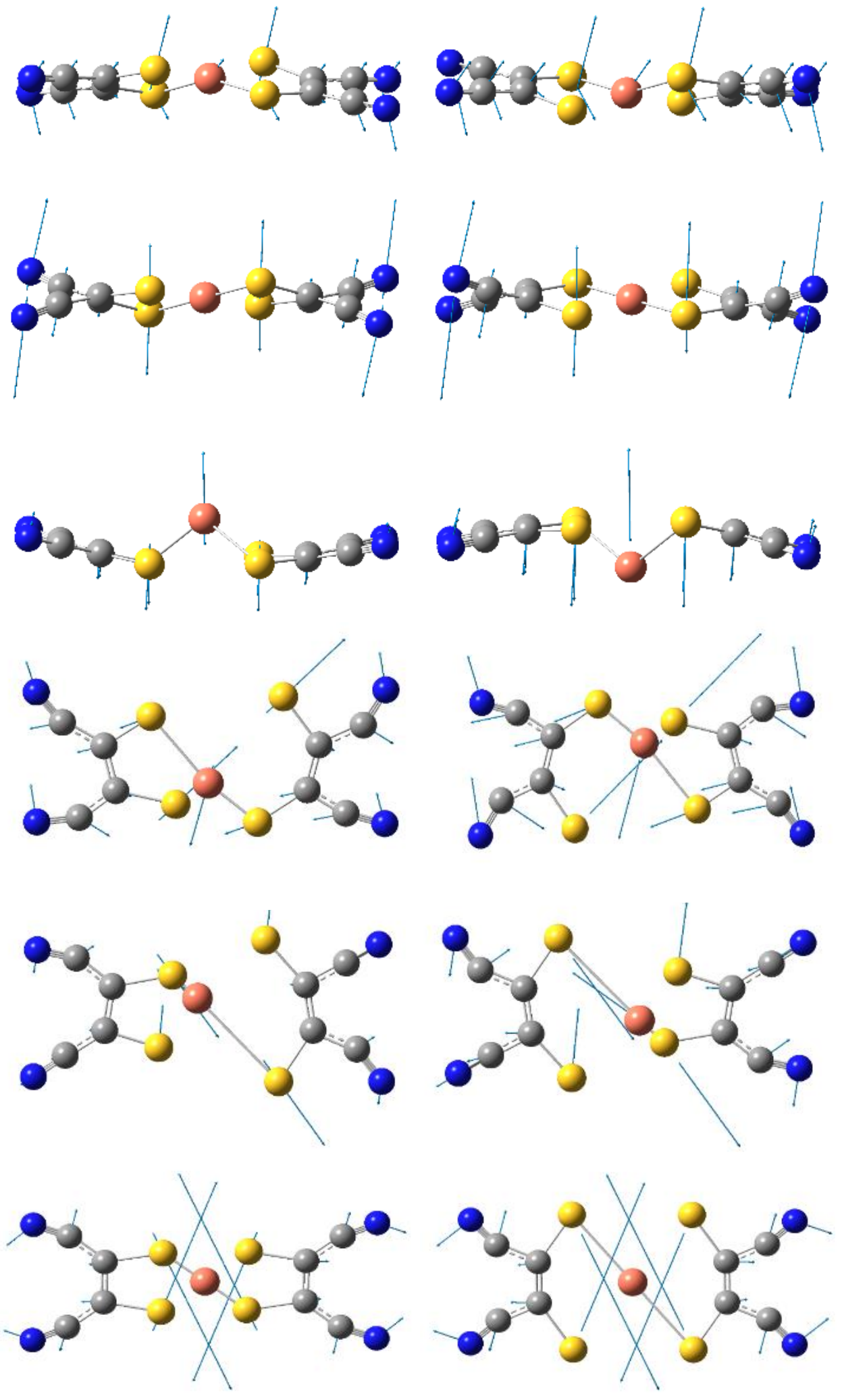

**Figure 4**. Displacement vectors (blue arrows) for some of the vibrational modes of **1**: 4 (first row), 12 (second row), 13 (third row), 23 (fourth row), 24 (fifth row), 25 (sixth row). These pictures have been taken for certain values of the generalized distortion coordinate: $Q_k$ = -1.0 Å (left); $Q_k$ = +1.0 Å (right).

For the $g_z$ calculations, the multi-configurational approach CASSCF/MS-CASPT2/RASSI-SO employing the software MOLCAS80 was selected as an *ab-initio* method. Among the first 25 vibrational modes of **1**, we found some that were symmetric ($g_z(-Q_k) = g_z(Q_k)$) and some which were antisymmetric ($g_z(-Q_k) = -g_z(Q_k)$); this property halved the number of calculations required for those modes. Such a feature should be exploited whenever we anticipate that the property *B* will be symmetric or antisymmetric respect to certain mode in order to reduce the computational cost.

Fig. 5 depicts the coupling constant $C_k$ of each one of the first 25 vibrational modes of **1**. There are five of them which clearly stand out from the rest: (a) modes 4 and 13, giving a positive coupling and (b) modes 12, 23 and 24, with a negative coupling. Note that the sign of $C_k$ is the sign of its second derivative. Thus, positive coupling constants tend to increase $g_z$ from its reference value at $T$ = 0 K, while negative ones tend to decrease it as temperature raises. The action of some of those modes that most intensely couple to the qubit is to distort the first coordination sphere of the magnetic ion, see Fig. 4. The existence of this kind of modes is practically unavoidable. Nevertheless, as stated previously, their effects could be minimized by encapsulating the metal.

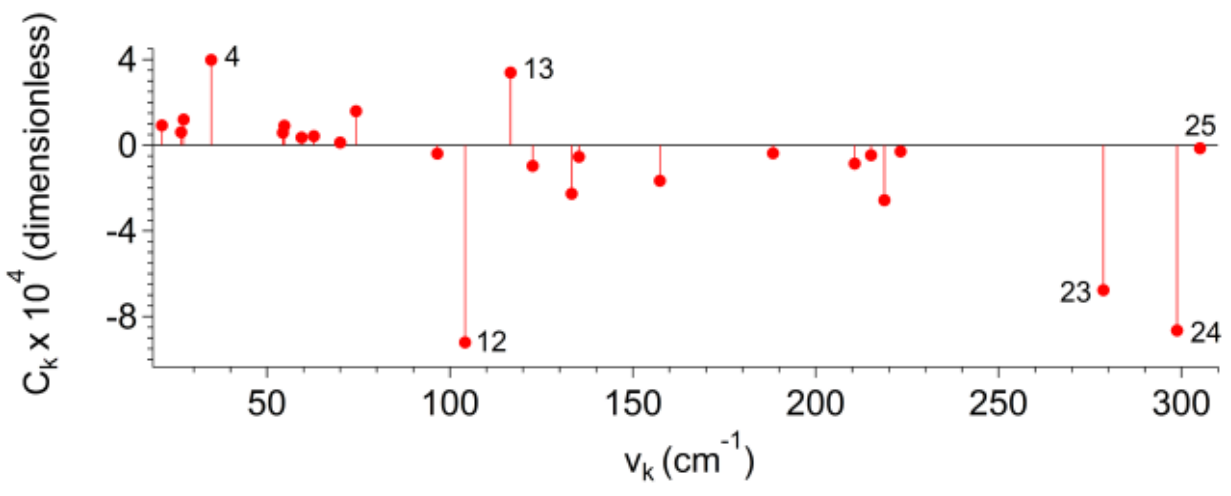

**Figure 5**. Spin-vibration coupling constants $C_k$ of the first 25 vibrational modes of **1** in Eq. (8).

From Figs. 4 and 5, the rationalization and prediction power of the above model can now be sketched. From chemical intuition, one would identify the most detrimental modes like those that mainly distort the coordination environment of the metal. This is true, for example, see modes 23 and 24 in figure 4, but not necessarily always. Indeed, we identify in **1** some modes which largely distort the coordination sphere but do not couple as much as expected, and modes with a significant coupling which hardly alter such a coordination sphere. The first case corresponds to the modes 13 and 25, while the modes 4 and 12 exemplify the second one. The mode 13 involves an important axial movement of the metal. Nevertheless, its coupling constant $C_k$ is unexpectedly smaller than that of the mode 4, and much smaller in magnitude than that of the mode 12, where in both cases the metal environment is quite less affected. The mode 25 is a breathing vibration, where the four sulphur atoms approach to and move away from the metal at once. Despite the significant distortion, its coupling constant $C_k$ is practically zero. Instead, the modes 23 and 24 produce a similar metal environment distortion, but their coupling constants are among the most important. This is a clear evidence of the crucial role of symmetric modes, favoured in high-symmetry molecules, in suppressing vibrational decoherence. Metal environment distortion is sometimes helped by the motion of mobile parts of the ligands, such as in the mode 21, see SI, which already displays a moderate coupling; or in the modes 23 and 24. This means that large ligands with mobile parts may also cause vibrational decoherence and should be engineered consequently. However, one could use ligands with mobile parts as well as they are small enough, like in **1**. Another advantage is the possibility of quantifying the coupling among a set of modes which seem to couple the same, giving the priority modes to engineer. For example, let us consider again the couple 4 and 12, where each mode distorts the metal environment pretty the same, but the individual coupling is considerably different. The capability of calculating the constant coupling is another advantage that can establish the right order in coupling within a pair of modes, an order that could result rather counter-intuitive *a priori*. For example, see the modes 12 and 13 again. As said previously, the mode 13 should couple more than the mode 12, as the former largely distort the metal environment more than the mode 12. However, mode 12 has a much higher coupling.

As said before, to obtain the spin energy modulation $B_k$ of each mode, one has to include not only its coupling strength but also its thermal population via its boson number. Fig. 6 gathers those individual modulations with temperature for each mode.

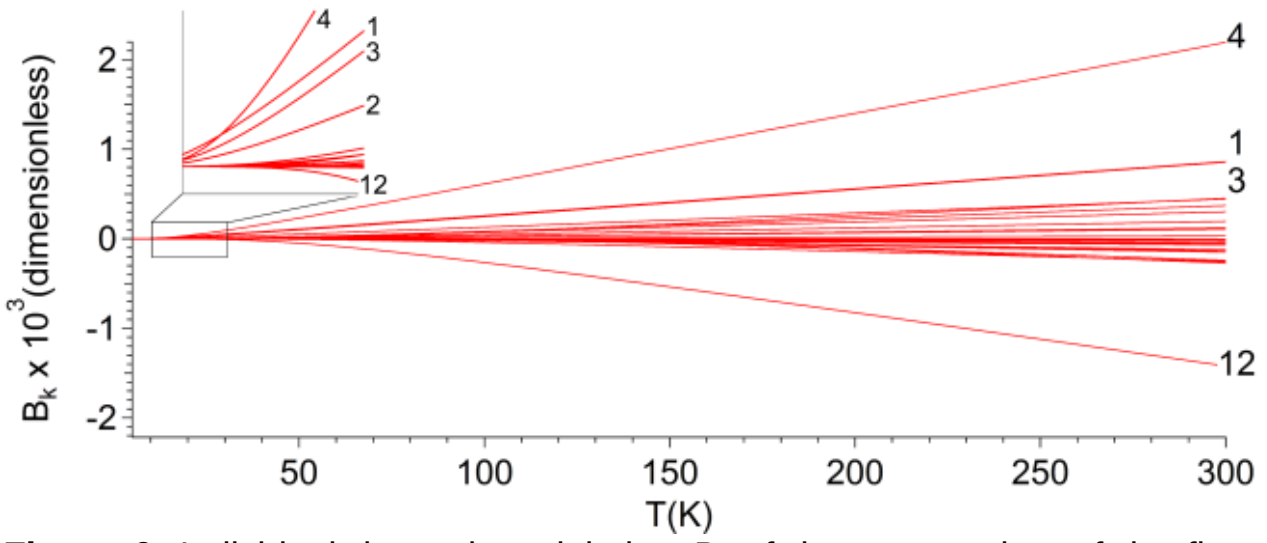

**Figure 6**: Individual thermal modulation $B_k$ of the expected $g_z$ of the first 25 vibrational modes of **1**. Inset: zoom-in at the range 10-30 *K*. Note that $B_1$ and $B_3$ have very similar thermal evolutions, as their slopes only differ in less than 1%.

At temperatures below 10-15 K, we see that the modulation is practically equal to zero. This observation is in concordance with the fact that vibrational decoherence does not have any important role at very low temperature. At higher temperatures, meaning $k_BT >> \nu_k$, each curve acquires a near linear behavior, since the boson number, Eq. (6), tends to a straight line with slope $k_B/\nu_k$ as temperature raises, see SI. Then, $B_k$ becomes a straight line with slope proportional to $(g_z)_{kk}/m_k\nu_k^2$, being $(g_z)_{kk}$ the second derivative of $g_z$ respect to $Q_k$. From 20 K on, where $k_BT$ is a sizeable fraction of $\nu_k$, modes begin to be populated appreciably and may contribute to the relaxation behavior. It is found that the twisting mode 4 clearly gives the largest positive contribution in the whole temperature range. This is possible because, besides its remarkable coupling, it is a low-frequency mode, $\nu_4$ = 34,76 cm$^{-1}$, so it becomes populated already from low T. On the contrary, the modes 23 and 24, despite its large coupling, are high-frequency modes, $\nu_{23}$ = 278,70 cm$^{-1}$ and $\nu_{24}$ = 298,81 cm$^{-1}$, so they become significantly populated only at high T and thus give a much weaker contribution. Something similar happens to the modes 1 and 3, whose coupling is smaller than that of some high-frequency modes, and the thermal evolutions of the former are above the ones of the latter. The mode 12 gives the largest negative modulation despite its coupling constant is similar to that of the modes 23 and 24. As mentioned above, the latter are high-frequency modes so they are irrelevant at room temperature, confirming the validity of our simplification of neglecting modes 26 and beyond.

One could think that if the thermal contributions of the modes cancelled each other it would result in a low vibrational decoherence at high T. Nevertheless, that may be a rather fallacious reasoning as the dynamics of each mode is generally different from the rest. Thus, those thermal contributions would not be necessarily globally cancelled given a temporal scale. This is the reason why vibrational modes should be individually engineered or, at least, looking for the best optimized global situation as said above.

Table 1 shows the experimental[13] and the statistical-averaged values of $g_z$ at low temperature (5 K) and room temperature (~294 K), as well as the corresponding relative thermal evolutions, $\Delta g_z(T)$, Eq. (9), in that range of temperatures:

$$\Delta g_z(T) = \frac{\overline{\langle g_z \rangle}(T) - \overline{\langle g_z \rangle}(5K)}{\overline{\langle g_z \rangle}(5K)} \cdot 1000 \tag{9}$$

**Table 1.** Statistical-averaged (up to the mode 25 of **1**) and experimental $g_z$ values at low and room temperatures together with the corresponding relative thermal evolutions, *Δg<sub>z</sub>(T)*, in the same temperature range.

| $g_z$ | 5 K | ~294 K | $\Delta g_z$(~294 K) |
|---|---|---|---|
| **Statistical Averaged** | 2.1380 | 2.1410 | + 1.403 ‰ |
| **Experimental ±0.0020** | 2.0932 | 2.0910 | [-2.959 ‰ ; + 0.861 ‰] |

As said before, only local modes are considered here, which means that we are not expecting to fully reproduce the $g_z$ thermal evolution. Nevertheless, that approximation offers an useful advantage. Indeed, by comparing the calculated and experimental evolutions, one can imagine the relative weight of other mechanisms of spin relaxation via Landé *g* factor, depending on the discrepancy. From Table 1, it is found that the calculated relative variation $\Delta g_z$ comes close to the experimental range [-2.959‰ ; +0.861‰]. Within the approximations made on this model, this should mean that local modes are responsible for the main contribution to vibrational decoherence via Landé *g* factor at high temperature,[9] while other mechanisms should have a weaker effect. In particular, lattice vibrations, which would confirm the lattice rigidity of **1** as stated in [13]. In a semi-empirical procedure, all those modulations $B_k$ to the expected $g_z$ would be usually reduced into one effective parameter, making it impossible to distinguish the magnitude of each one of them. Another simplification could be to consider that vibrational decoherence is governed by only one mode. While this may work in some cases, the case study **1** shows us that the vibrational nature of molecules can be complex enough to discard that hypothesis.

At the 100 K experimental geometry, the calculated value of $g_z$ following the *ab-initio* methodology was 2.1233, while its experimental value at 5 K is 2.0932 ± 0.0020. A procedure (not yet published) adapted from that reported in [22] followed by a DDCI (Difference Dedicated Configuration Interaction Method) evaluation of the spin-free spectrum gave us a more accurate value of 2.0990, but its cost is prohibitive to perform the required number of calculations and thus we followed the original MS-CASPT2 procedure. Nevertheless, that more sophisticated calculation does validate our approach as it shows that disposing of accurate parameters is possible, so the performance of the model is simply reduced to an available quality matter of those parameters. We could not calculate even a more accurate DDCI-like $g_z$ value since experimental geometries are not usually determined at temperatures as low as 5 K, a limitation already pointed out in [7(a)].

## Conclusions

Both in Molecular Spin Qubits and in Single Ion Magnets, the high-temperature relaxation behavior is governed by the interaction between vibrations and spin excitations. Therefore, it is necessary to clarify how crystal and molecular structures determine that spin-vibration interaction, firstly in order to rationalize the latest experimental advances in these two fields but also with the goal of designing new molecular complexes that display slow relaxation at high temperatures. We have presented here a straightforward scheme to quantify, from first principles and for any given structure, the interaction between a spin excitation and all different vibrations. The scheme starts by the theoretical determination of the vibrational modes in terms of frequencies, reduced masses and displacement vectors; these displacement vectors are subsequently used in a series of calculations aimed to determine how vibrations affect the magnetic energy level scheme. In particular, they allow determining the variation of the qubit energy with vibrational distortions, either directly or indirectly. In our case this happens indirectly, in the form of the Landé $g_z$ factor. In the final step, equations (3)-(8) derived herein can be used to estimate a spin-vibration interaction constant *C* for every individual vibration. To showcase the procedure we employ the complex $[Cu(mnt)_2]^{2-}$. The theoretical framework allowed by this method is helpful in understanding recent experimental successes and guide future work: we can now quantitatively identify the vibrations that most intensely affect the energy of the spin qubit, e.g. by altering the magnetic anisotropy, zero-field splitting or quantum tunneling gap of the magnetic complex. These vibrations will mediate in the energy dissipation pathway between the spin energy levels and the thermal bath and therefore understanding and controlling them can be instrumental in understanding and controlling the decoherence process. Generally speaking, we find that either a high vibrational frequency or a low value of the spin-vibration interaction constant *C* are enough to quench the interaction of the spin qubit energy with a given vibrational mode. Moreover we find that, practically, the focus needs to be put on the vibrations with the lowest vibrational frequency, and among those the ones with high values of *C*. In particular, we can now quantify the benefit of systems where any movement of the ligands that affects the qubit energy corresponds to a high frequency mode, as it happens in porphyrines or phtalocyanines, where twisting vibrations such as those found here would be of very high frequency. Equally, we can now properly value the importance of avoiding the movement of hydrogen in the critical vibrations, since light atoms govern the reduced mass of the vibration, which is inversely proportional to the spin-vibration coupling constant *C*; indeed, when hydrogen is involved in vibrations affecting the spin energy levels, deuteration or fluorination is expected to have a dramatic effect on the vibrational decoherence time. This might be another factor for the success in cases where the qubit ion is encapsulated in a rigid carbon cage as in $N@C_{60}$. Finally, one needs to note that the methodology presented here is rather general, since it can incorporate both lattice and molecular vibrations; furthermore it is extensible to include anharmonicity and higher order terms in the Taylor expansion describing the qubit energy dependency with vibrational coordinates. Studying larger molecular complexes would also be feasible by using parametrical methods, such as the one implemented in the software SIMPRE.[23] Since the methodology can be implemented using standard tools of computational chemistry available to most chemists, such as the Gaussian and MOLCAS suites, it should be widely applicable in the design of more resilient molecular spin qubits or single ion magnets that maintain their spin state at high temperature.

## Acknowledgements

We thank Prs. Nathalie Guihéry and Jean-Paul Malrieu for fruitful discussions. The research reported here was supported here by the spanish MINECO (grants MAT2015-68204-R, CTQ2015-64486-R, and Excellence Unit María de Maeztu MDM-2015-0538), the European Union (ERC grant DECRESIM, and COST 15128 Molecular Spintronics Project) and the Generalitat Valenciana (Prometeo and ISIC-Nano Programs of Excellence). A.G.-A. thanks the Spanish MINECO for a Ramón y Cajal

Fellowship. L.E.-M. acknowledges the Generalitat Valenciana for a VALi+D predoctoral contract. N.S. thanks Université Toulouse III Paul Sabatier and CNRS for fundings. This work was performed using HPC resources from CALMIP (Grant 2015-1144).

**Keywords:** Molecular Magnetism • Molecular Spin Qubit • Coordination Chemistry • Computational Chemistry • Spin-Phonon Coupling

**Supplementary Information**

**Role of vibrations on decoherence in molecular spin qubits: The case of $Cu(mnt)_2^{2-}$**

Luis Escalera-Moreno[a], Nicolas Suaud[b], Alejandro Gaita-Ariño[a] and Eugenio Coronado[a]

1. Details on geometry optimization and vibrational spectrum calculation of **1** ($Cu(mnt)_2^{2-}$, $mnt^{2-}$ = 1,2-dicyanoethylene-1,2-dithiolate).

In the case of **1**, we used UB3LYP as a functional, 6-31G** as a basis set for each atom, and the package Gaussian09. In addition, we optimized its geometry not in vacuum but by additionally considering a portion of its environment. In order for this calculation not to be too expensive, we applied two approximations: (a) the counter-ions of the selected environment were kept frozen during the optimization so that only the complex was optimized, and (b) all the atoms of such an environment were changed by helium atoms before running the optimization.

This approach allowed us to get an optimized geometry very close to the experimental one at low temperature (100 K), where the relative deviations of the Cu-S bond lengths and S-Cu-S bond angles from the experimental values were around 3% and 1%, respectively. A crucial point should be noted here. We need a starting geometry with no temperature in order to then include in it the molecular vibrational population by means of a Boltzmann statistic. In other words, when optimizing the complex, we are looking for the geometry corresponding to the bottom of its potential energy surface. Because of the zero-point energy, there are no experimental geometries at the bottom but it is reasonable to think that the lower the temperature is the closer the experimental geometry will be to a virtual experimental geometry at the bottom (the maximum of probability is located at the bottom for the ground vibrational energy level n=0). This is the reason why we compare our optimized geometry to the experimental one just at low temperature. As a matter of fact, it should be always compared to an experimental geometry at a temperature as low as possible, never at high temperature (except if one wants to include temperature in an effective way in the geometry). An optimized geometry has no temperature, while an experimental one always includes it. Those geometries are different from a conceptual point of view, getting closer quantitatively only at low temperature. As magnetic anisotropy strongly depends on the environment of the metal, anisotropy parameters like the Landé g factor may also be a closeness measure between the optimized and the experimental geometries. In this sense, the calculated $g_z$ at our optimized geometry is $(g_z)_e$ = 2.1390, satisfactorily close to the experimental value of 2.0932 ± 0.0020 at 5K, indicating that optimized geometry may be close to the bottom as desired.

2. Details on Wave-Function Based Theory calculations to evaluate the magnetic anisotropy of **1**.

Given a geometry, the evaluation of the electron Landé *g* factors is performed following the procedure detailed in [S1]. The wave functions of the seven $Cu^{2+}(d^9)$ lowest energy electron spin doublets are evaluated at the CASSCF level by using an active space that considers 13 electrons in 12 molecular orbitals (two copper-ligand σ-bonding orbitals, the five copper 3d magnetic orbitals and five additional virtual 3d molecular orbitals). The weights of those seven roots are 5 3 3 3 3 1 1, respectively. A second order perturbative dynamical correction to the energy is then obtained through a MS-CASPT2 (multi-state) calculation with an imaginary shift of 0.04. Subsequently, spin-orbit coupling is taken into account by using the RASSI-SO module of the MOLCAS80 package.[S2] This methodology was validated for square-planar copper complexes in [S3]. The contracted ANO-RCC basis sets 6s4p3d1f for Cu, 5s4p1d for S and 4s3p1d for C and N were used.

We assume that the direction of the magnetic field, the z-direction, is that of the easy axis of the optimised geometry. Thus, at the equilibrium geometry, if $g_1$, $g_2$ and $g_3$ are the Landé g factors calculated in the three spatial directions, we take as a $g_z$ the highest g of those three values. When the complex vibrates, the easy axis changes its direction, not coinciding thus with the direction of the magnetic field. For each one of the distorted geometries, MOLCAS calculates, as said previously, the g factor of its easy axis in particular. However, what we really need is the g value acting in the magnetic field direction, not in the direction of the easy axis of that distorted geometry. To get that correct g value, this means that the g-tensor G' of each distorted geometry needs to be expressed in the basis set that diagonalizes the equilibrium geometry g-tensor G. If R is the invertible matrix that diagonalizes G, i.e., $R^{-1}GR$ = Diag, the desired representation of G' is $R^{-1}G'R$, which is not necessarily a diagonal matrix. The $g_z$ value that we save for that distorted geometry is just the highest value of the main diagonal of $R^{-1}G'R$. Note that, in another software, the diagonalization could be given by $RGR^{-1}$ = Diag and then thus correction would be $RG'R^{-1}$.

## **3.** Details on the generation of the distorted geometries for each vibrational mode of **1**.

The Landé g factor calculations are performed at the DFT equilibrium geometry and at certain number of distorted geometries, which are obtained by applying, to this equilibrium geometry, the displacement vector of each vibrational mode. This means that for each mode k, those distorted geometries are generated by giving suited values to its one-dimensional generalized distortion coordinate $Q_k$. The equilibrium geometry is recovered when $Q_k = 0$ for every k.

In order to properly generate distorted geometries, the minimum distortion between any consecutive pair of them has to be at least above the X-ray experimental error to produce a significant distortion. Let $v_1$ and $v_2$ be the 3P-dimensional cartesian coordinate vectors of two distorted geometries of the mode k. Then, for some value of $Q_k$, the 1D distortion coordinate, they are related as $v_2 = v_1 + Q_k \cdot v_k$, where $v_k$ is the normalised displacement vector, $|v_k| = 1$. Thus $|Q_k| = |v_2 - v_1|$. We need to choose $Q_k^{min}$ such that at least one component of the vector $v_2 - v_1$ is at least above its experimental error for the distortion to be significant. So, let v be the 3P-dimensional vector whose components are the cartesian (a.k.a. orthogonal) experimental errors of each coordinate of each atom. The desired criterion is easily met by taking $|Q_k^{min}| = |v|$. As the experimental error is obviously the same for each vibrational mode, the chosen $|Q_k^{min}|$ does not depend on k and thus it is the same for all of them, $|Q^{min}| = |v|$. For our system, we have $|Q^{min}|$ = 0.022 Å. Therefore, we will generate distorted geometries giving $Q_k$ integer multiples of 0.022 Å. If the experimental error of each coordinate t (t = x,y,z) is given in fractional coordinates, it is quickly converted to cartesian coordinates by multiplying it by the corresponding cell parameter a (t = x), b (t = y) or c (t = z). As the distortion coordinate $Q_k$ cannot take arbitrarily large values, because it would generate geometries beyond the harmonic regime, an upper bound to $|Q_k|$ must be selected. We take the smallest natural s such that 0.022s is just above the classical limit of the first excited vibrational level n = 1. This way, we make sure that distortions will be small enough to be inside the harmonic regime. The distorted geometries are then the ones generated with $Q_k$ = 0.022j, j = -s, …, -1, 1, …, s. Since the derivative is a local concept, we only need the fitting polynomial, to extract its second derivative, in a small neighbourhood of $Q_k$ = 0. Thus, giving $|Q_k|$ a small upper bound is not a problem at all.

4. Table S1. List of the harmonic frequencies "v" and reduced masses "m" of the first 25 vibrational modes of **1**. u.m.a.q. = chemical atomic mass units, the twelfth part of the mass of an atom of the isotope $^{12}C$.

| **Mode** | **1** | **2** | **3** | **4** | **5** |
|---|---|---|---|---|---|
| v ($cm^{-1}$) | 21.2781 | 26.5579 | 27.2106 | 34.7649 | 54.3558 |
| m (u.m.a.q.) | 20.5084 | 20.9844 | 14.5025 | 23.1318 | 26.5854 |
| **Mode** | **6** | **7** | **8** | **9** | **10** |
| v ($cm^{-1}$) | 54.7769 | 59.4704 | 62.8441 | 70.0300 | 74.3575 |
| m (u.m.a.q.) | 16.2118 | 17.3111 | 16.3245 | 15.3684 | 22.2152 |
| **Mode** | **11** | **12** | **13** | **14** | **15** |
| v ($cm^{-1}$) | 96.5870 | 104.1581 | 116.6254 | 122.6491 | 133.2913 |
| m (u.m.a.q.) | 16.1088 | 19.3045 | 35.3246 | 15.3963 | 16.8291 |
| **Mode** | **16** | **17** | **18** | **19** | **20** |
| v ($cm^{-1}$) | 135.3013 | 157.4077 | 188.3768 | 210.6606 | 215.1370 |
| m (u.m.a.q.) | 14.2572 | 16.1081 | 19.5370 | 32.5990 | 13.5001 |
| **Mode** | **21** | **22** | **23** | **24** | **25** |
| v ($cm^{-1}$) | 218.8099 | 223.2432 | 278.6986 | 298.8124 | 305.0957 |
| m (u.m.a.q.) | 20.6210 | 13.5662 | 29.0062 | 32.2432 | 30.2748 |

5. Table S2. Extreme values $Q_k^{max}$ of the distortion coordinates $Q_k$ for the first 25 vibrational modes of **1**. The smallest value of $Q_k$ is ± 0.022 Å. Any intermediate value of $Q_k$, including $Q_k^{max}$, is an integer multiple of 0.022 Å as discussed in 3. $Q_k^{max}$ is the smallest integer multiple of 0.022 Å above the turning point of the first excited harmonic vibrational state n = 1 of each mode.

| **Mode** | **1** | **2** | **3** | **4** | **5** |
|---|---|---|---|---|---|
| $Q_k^{max}$ (Å) | ± 0.484 | ± 0.440 | ± 0.528 | ± 0.374 | ± 0.286 |
| **Mode** | **6** | **7** | **8** | **9** | **10** |
| $Q_k^{max}$ (Å) | ± 0.352 | ± 0.330 | ± 0.330 | ± 0.308 | ± 0.264 |
| **Mode** | **11** | **12** | **13** | **14** | **15** |
| $Q_k^{max}$ (Å) | ± 0.264 | ± 0.242 | ± 0.176 | ± 0.242 | ± 0.220 |
| **Mode** | **16** | **17** | **18** | **19** | **20** |
| $Q_k^{max}$ (Å) | ± 0.242 | ± 0.220 | ± 0.176 | ± 0.132 | ± 0.198 |
| **Mode** | **21** | **22** | **23** | **24** | **25** |
| $Q_k^{max}$ (Å) | ± 0.154 | ± 0.198 | ± 0.132 | ± 0.110 | ± 0.110 |

6. Figures S1 – S20. Calculated, and subsequently corrected (as explained in 2), $g_z$ factors as a function of the distortion coordinate $Q_k$ along with the fitting polynomial for the first 25 vibrational modes of **1**.

Another point to properly calculate the second derivatives is to avoid, as much as possible, numerical errors. In particular, the g factor rounding to the fourth decimal as done in MOLCAS. Due to that rounding, the printed g factors are just approximations to the real values. As a consequence, the fitting polynomial does not necessarily must match those rounded values and, thus, the "*closest regression coefficient $R^2$ to 1*" criterion is not a reliable rule to select that polynomial. Assuming that the real curve $g_z = g_z(Q_k)$ is a smooth function of $Q_k$, the fitting polynomial is taken as the polynomial which gives the smoothest trend with the lowest possible degree. Although this criterion may seem rather subjective, it has been tested that changing the selected degree around one or two unities produces a completely negligible perturbation in the second derivatives. The rounding limitation is a crucial factor that should be overcome to improve the quality of the parameters herein calculated.

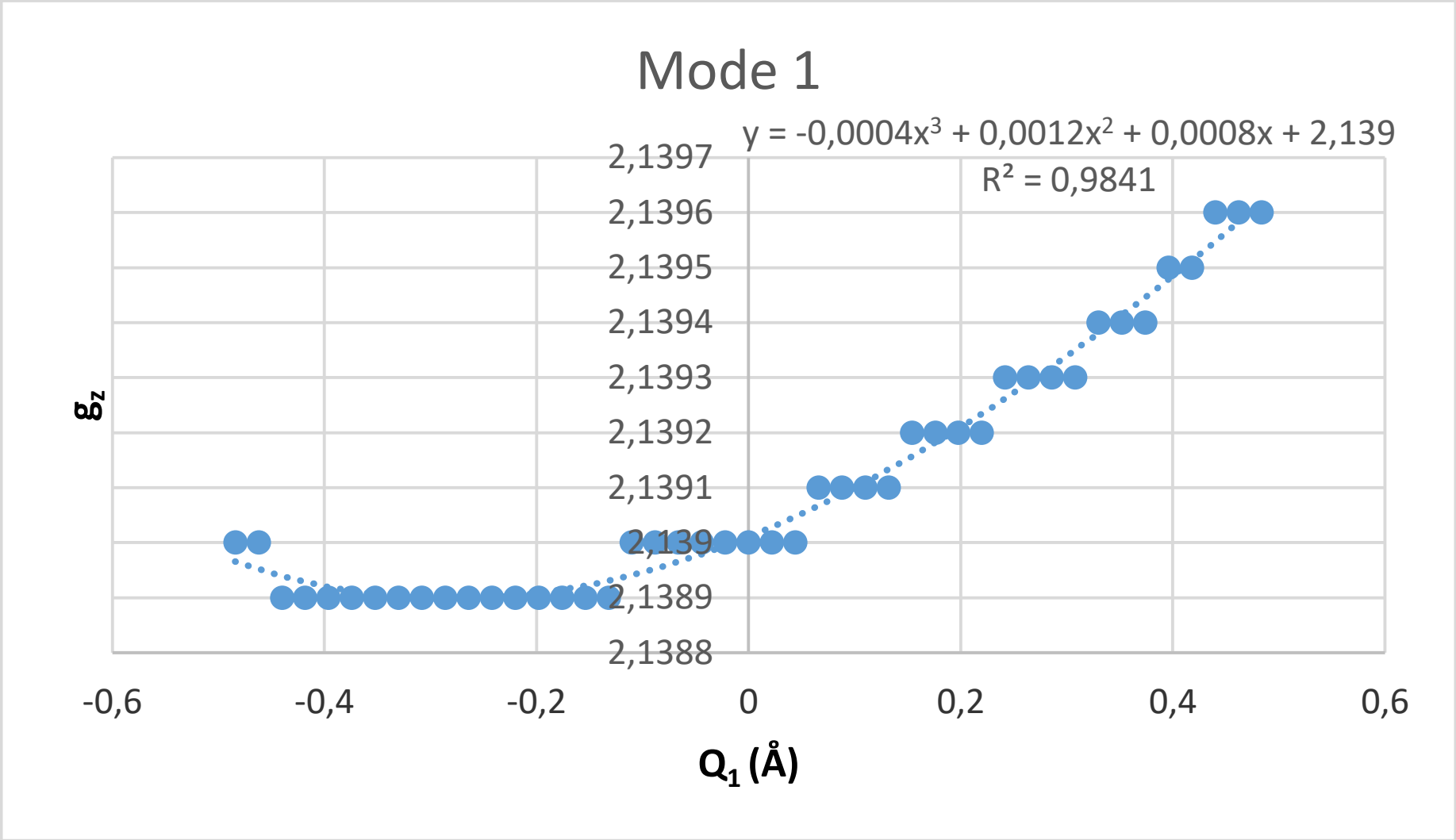

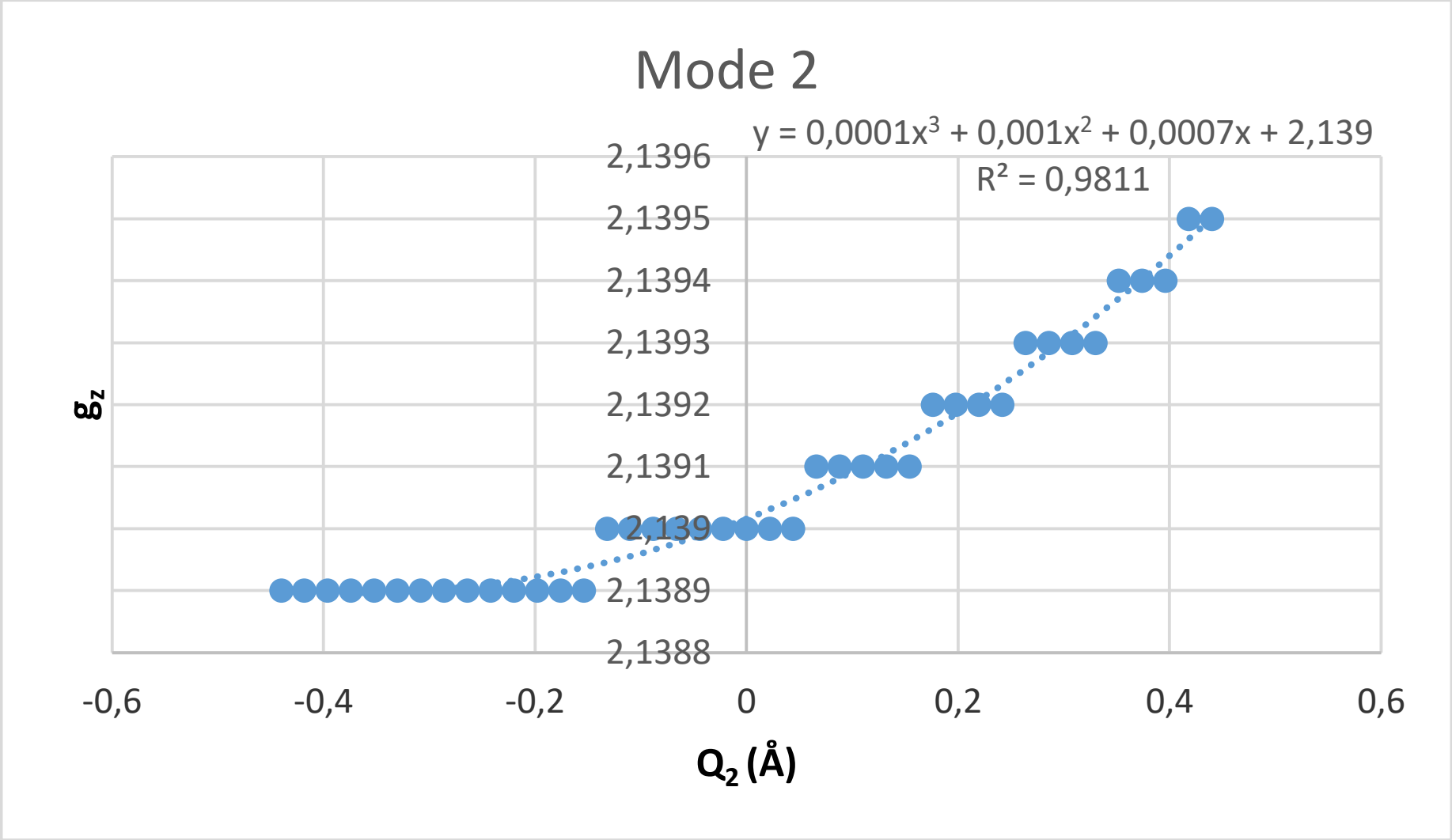

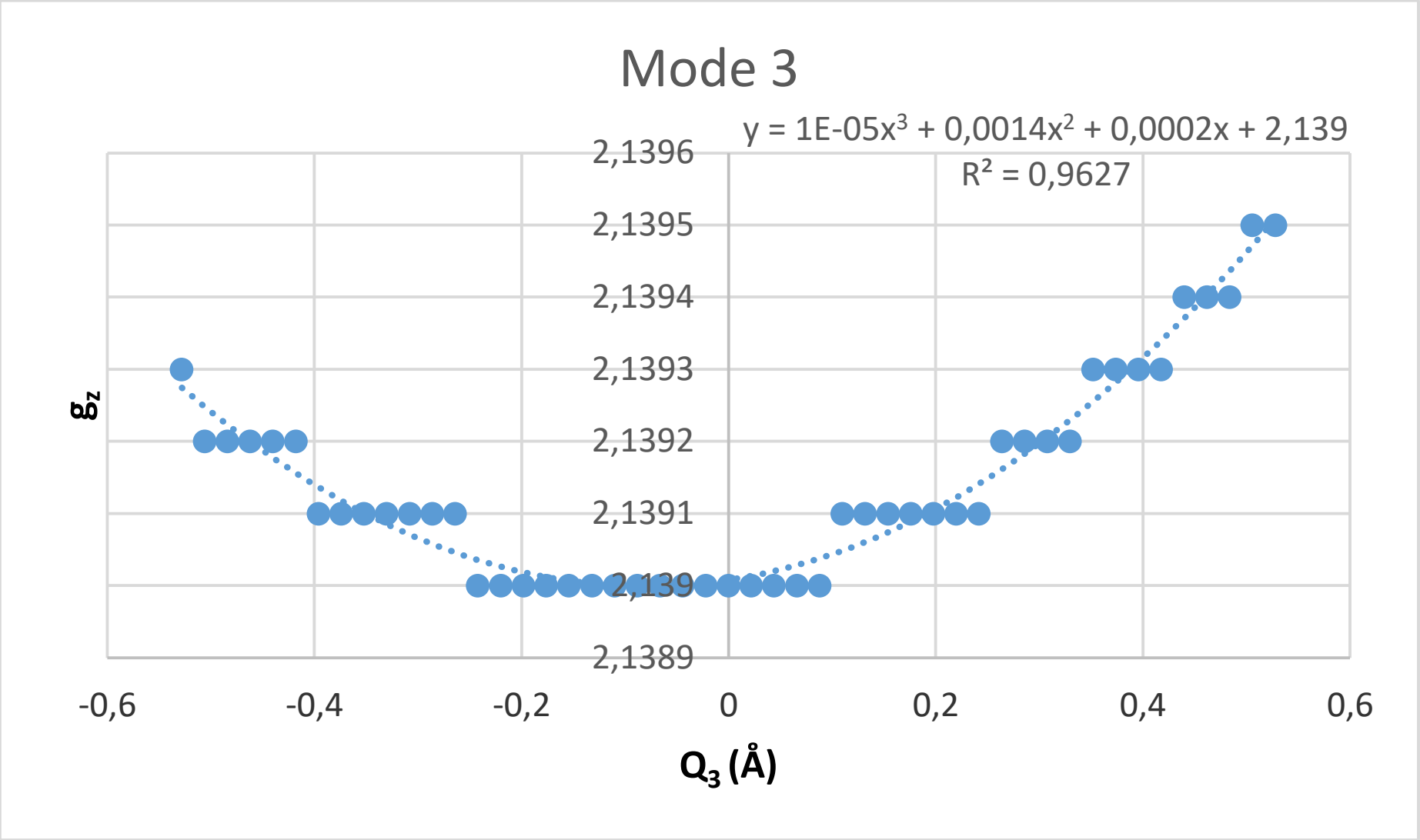
Mode 3
y = 1E-05x³ + 0,0014x² + 0,0002x + 2,139
R² = 0,9627
2,1396
2,1395
2,1394
2,1393
2,1392
2,1391
2,139
2,1389
-0,6
-0,4
-0,2
0
0,2
0,4
0,6
gz
Q3 (Å)

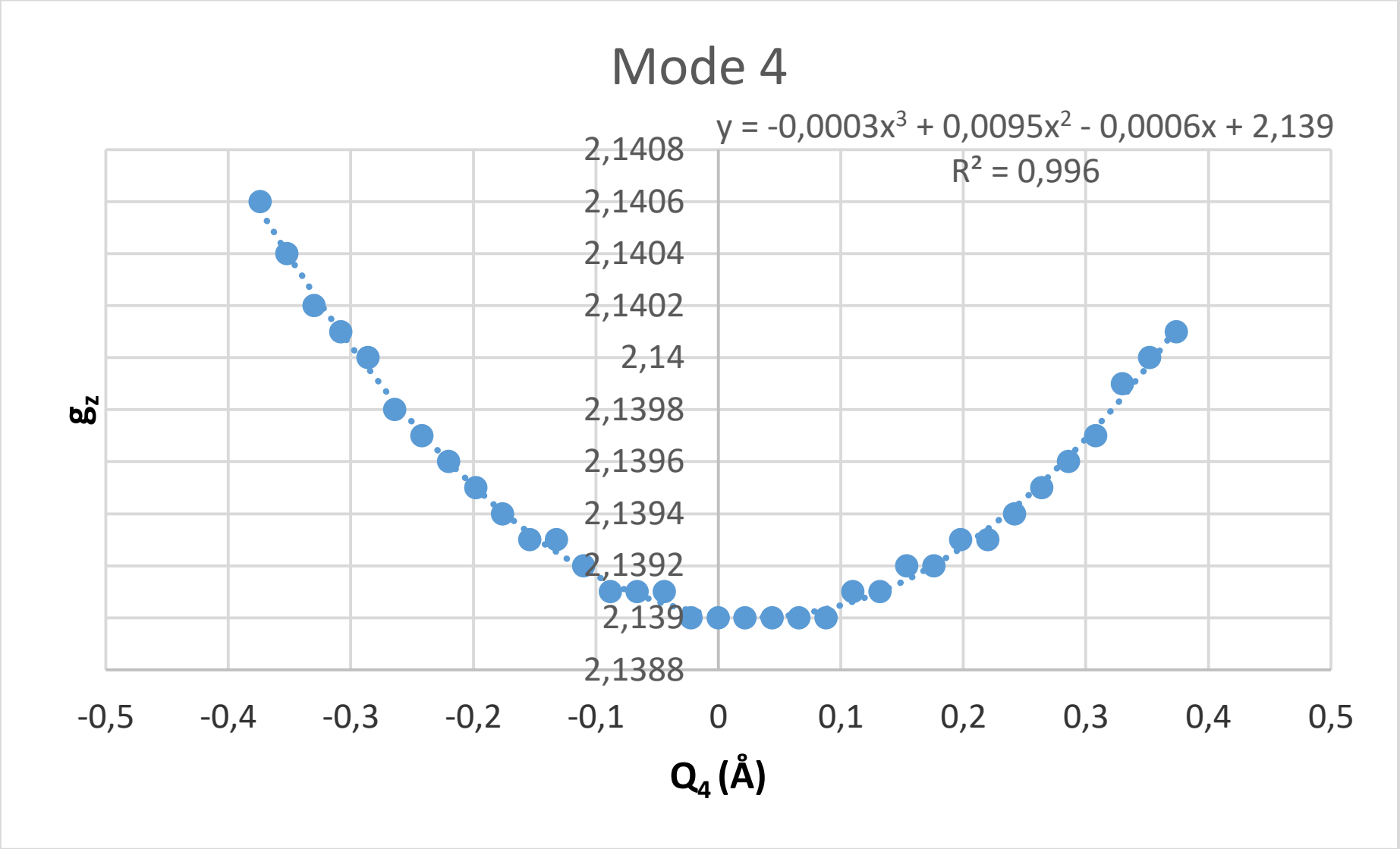
Mode 4
y = -0,0003x³ + 0,0095x² - 0,0006x + 2,139
R² = 0,996
2,1408
2,1406
2,1404
2,1402
2,14
2,1398
2,1396
2,1394
2,1392
2,139
2,1388
-0,5
-0,4
-0,3
-0,2
-0,1
0
0,1
0,2
0,3
0,4
0,5
gz
Q4 (Å)

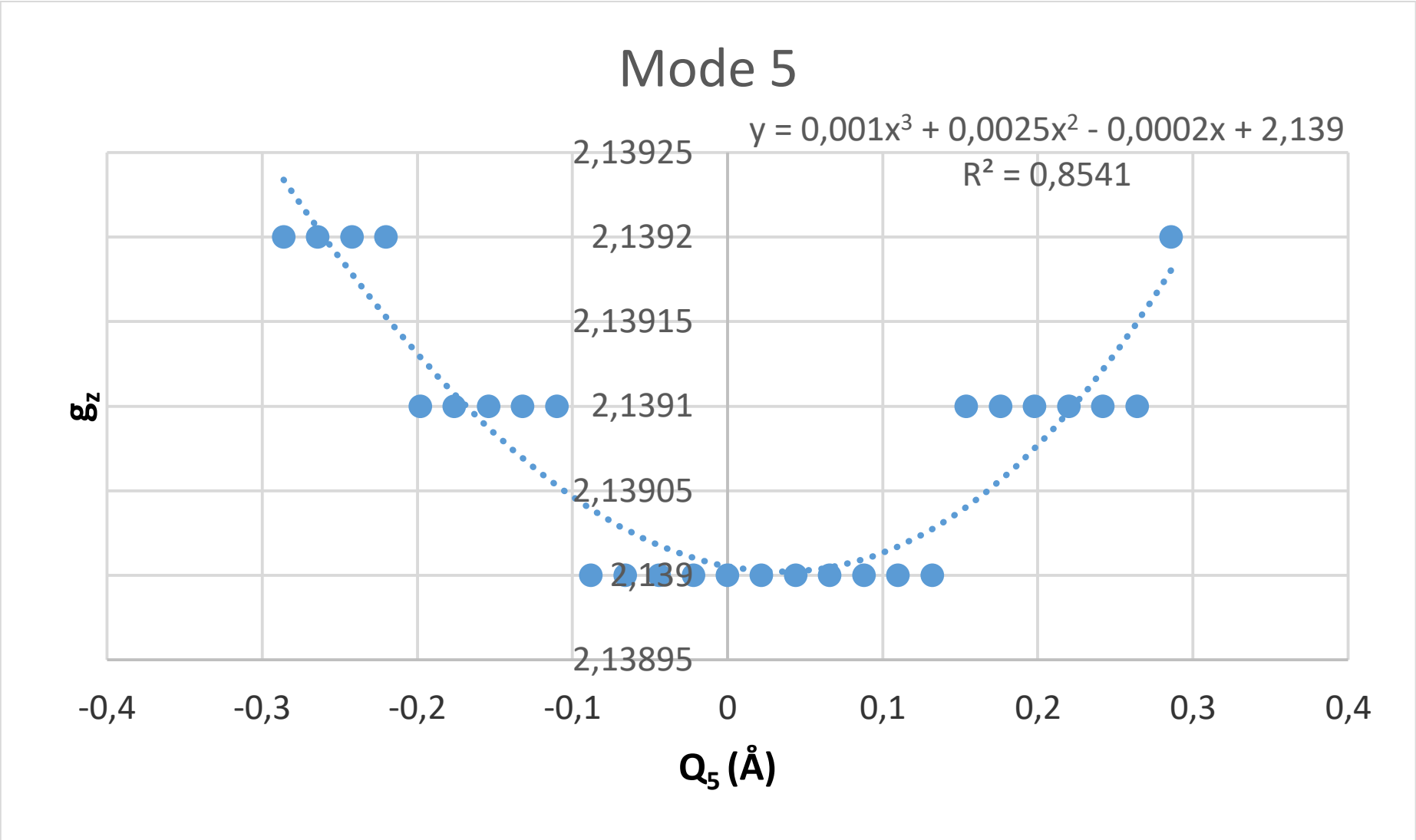
Mode 5
y = 0,001x³ + 0,0025x² - 0,0002x + 2,139
R² = 0,8541
2,13925
2,1392
2,13915
2,1391
2,13905
2,139
2,13895
-0,4
-0,3
-0,2
-0,1
0
0,1
0,2
0,3
0,4
gz
Q5 (Å)

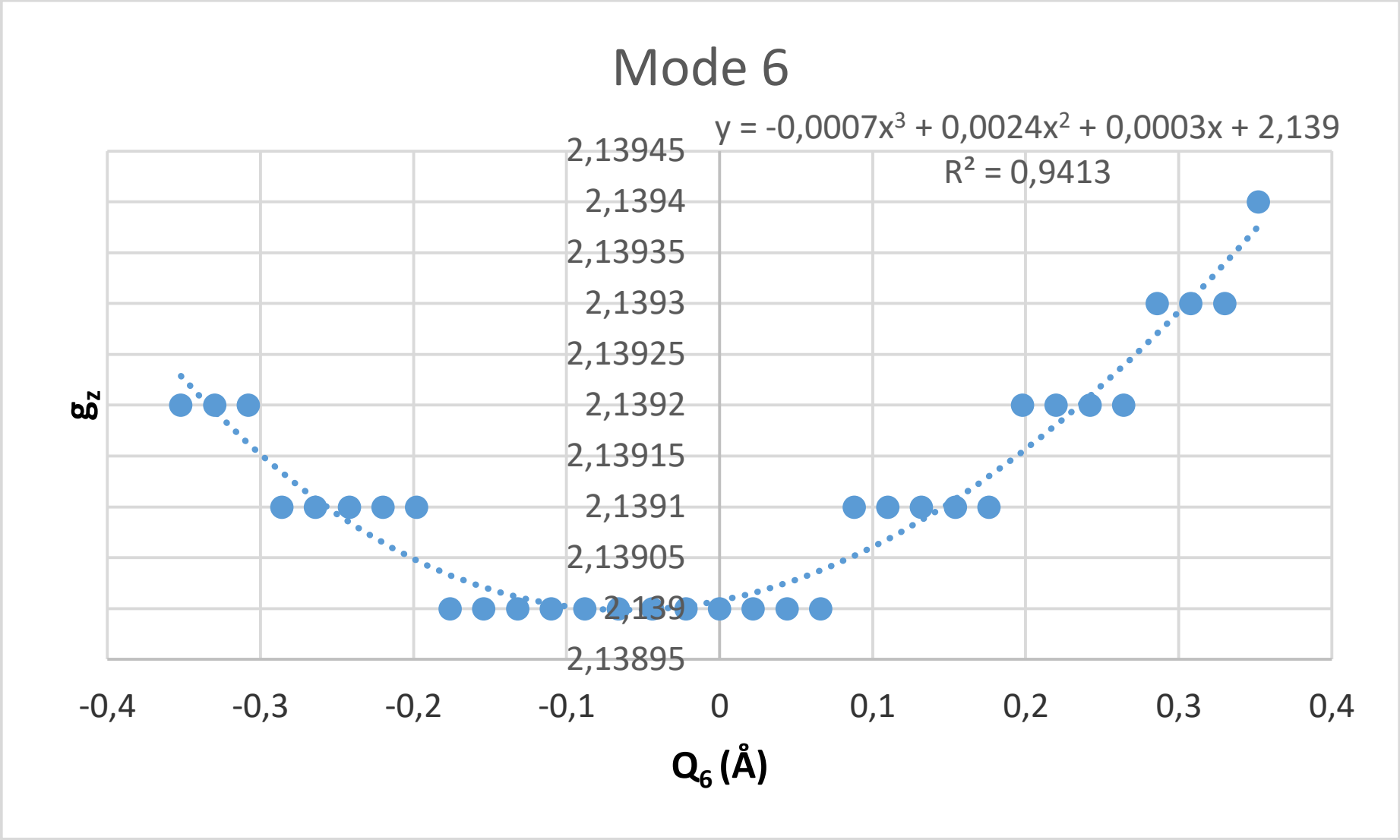
Mode 6
y = -0,0007x³ + 0,0024x² + 0,0003x + 2,139
R² = 0,9413
2,13945
2,1394
2,13935
2,1393
2,13925
2,1392
2,13915
2,1391
2,13905
2,139
2,13895
gz
-0,4
-0,3
-0,2
-0,1
0
0,1
0,2
0,3
0,4
Q6 (Å)

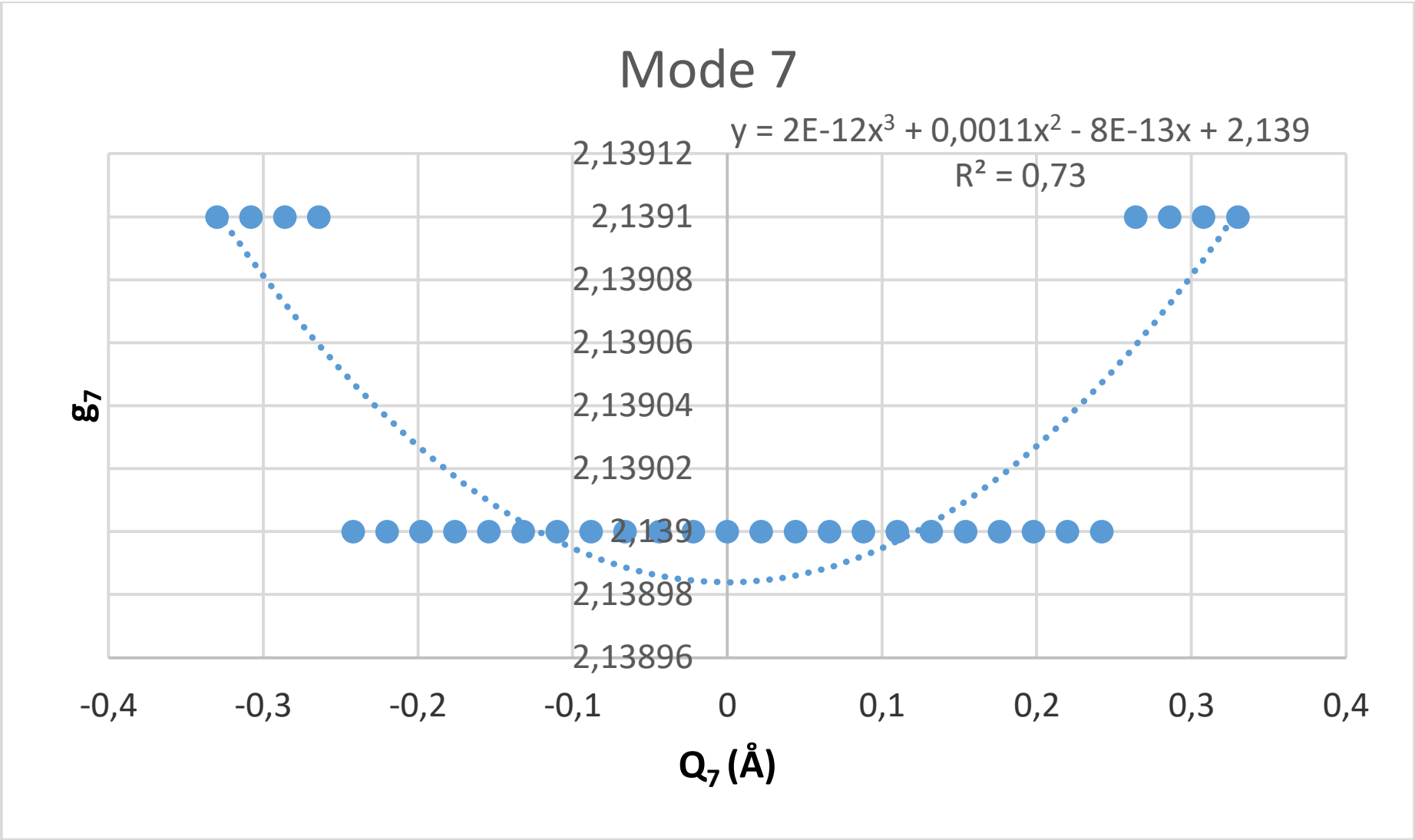
Mode 7
y = 2E-12x³ + 0,0011x² - 8E-13x + 2,139
R² = 0,73
2,13912
2,1391
2,13908
2,13906
2,13904
2,13902
2,139
2,13898
2,13896
g7
-0,4
-0,3
-0,2
-0,1
0
0,1
0,2
0,3
0,4
Q7 (Å)

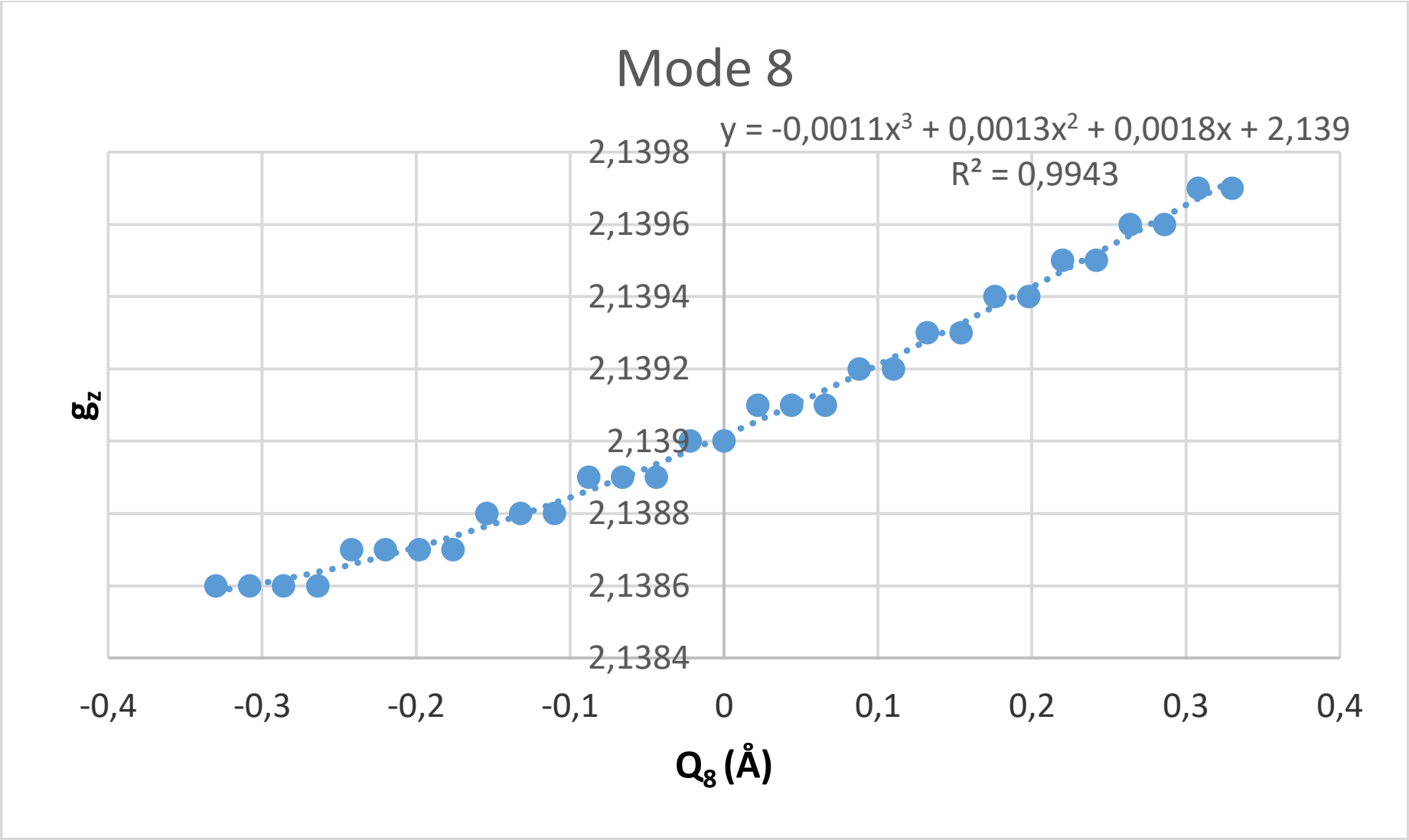
Mode 8
y = -0,0011x³ + 0,0013x² + 0,0018x + 2,139
R² = 0,9943
2,1398
2,1396
2,1394
2,1392
2,139
2,1388
2,1386
2,1384
gz
-0,4
-0,3
-0,2
-0,1
0
0,1
0,2
0,3
0,4
Q8 (Å)

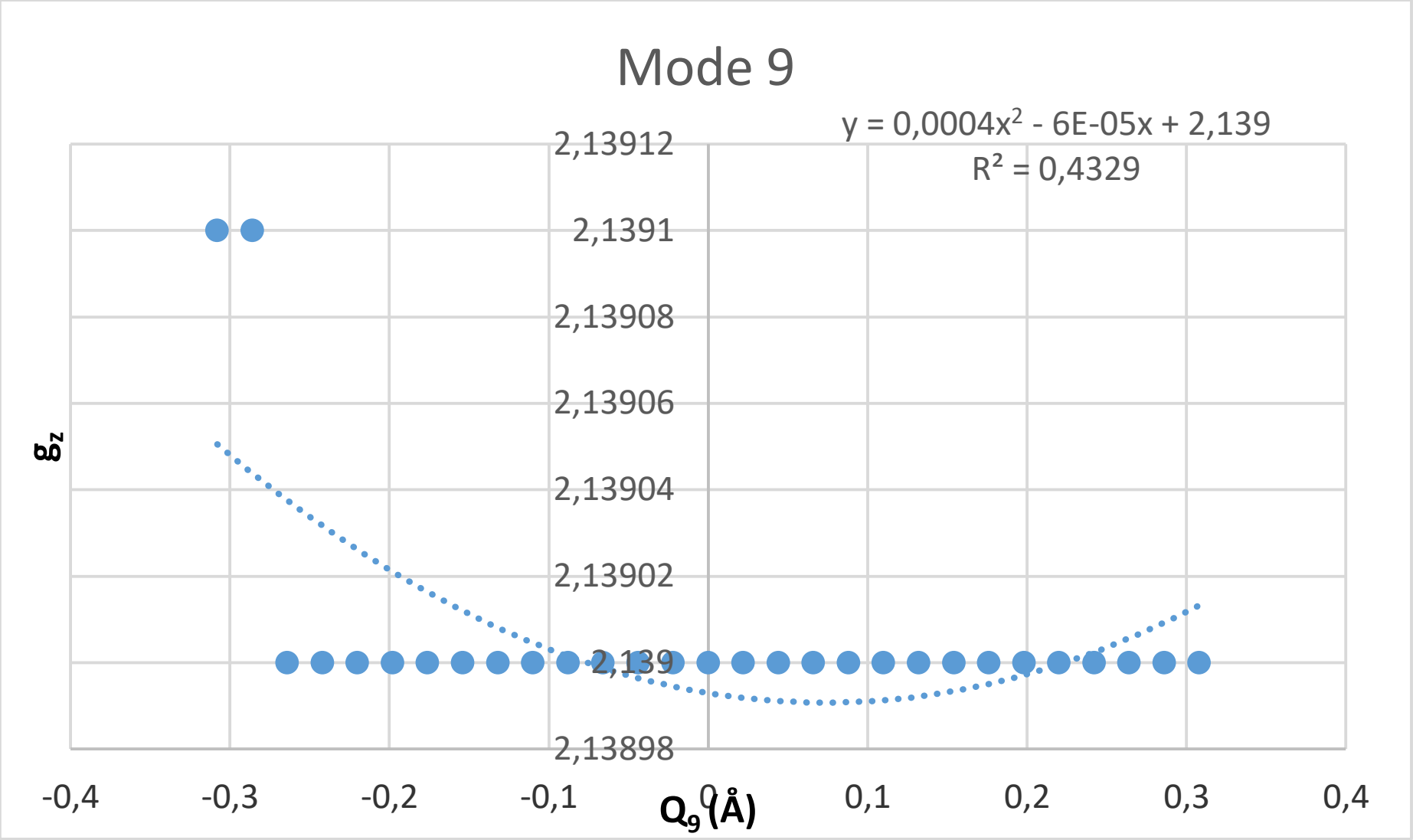

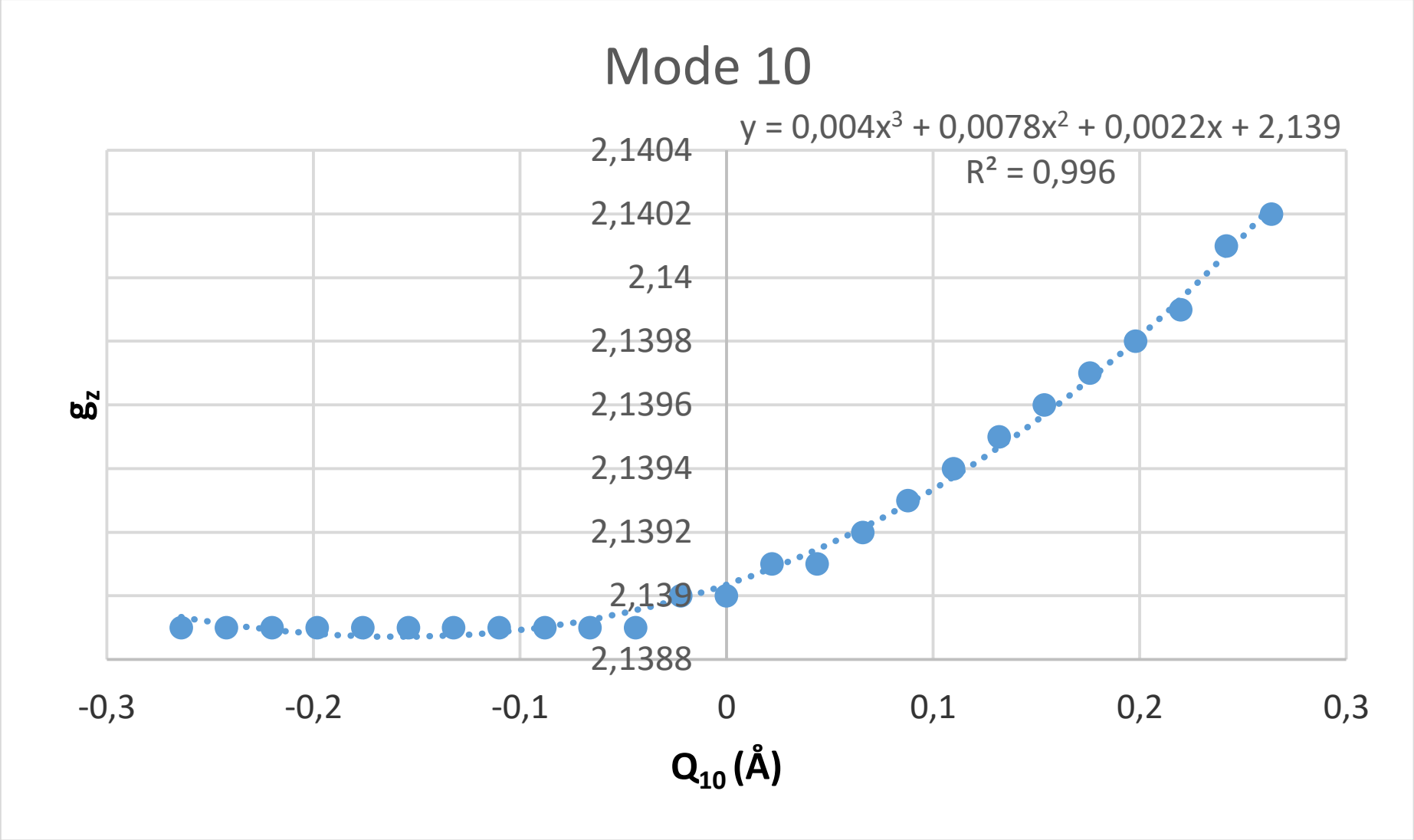
Mode 10
y = 0,004x³ + 0,0078x² + 0,0022x + 2,139
R² = 0,996
2,1404
2,1402
2,14
2,1398
2,1396
2,1394
2,1392
2,139
2,1388
gz
-0,3
-0,2
-0,1
0
0,1
0,2
0,3
Q10 (Å)

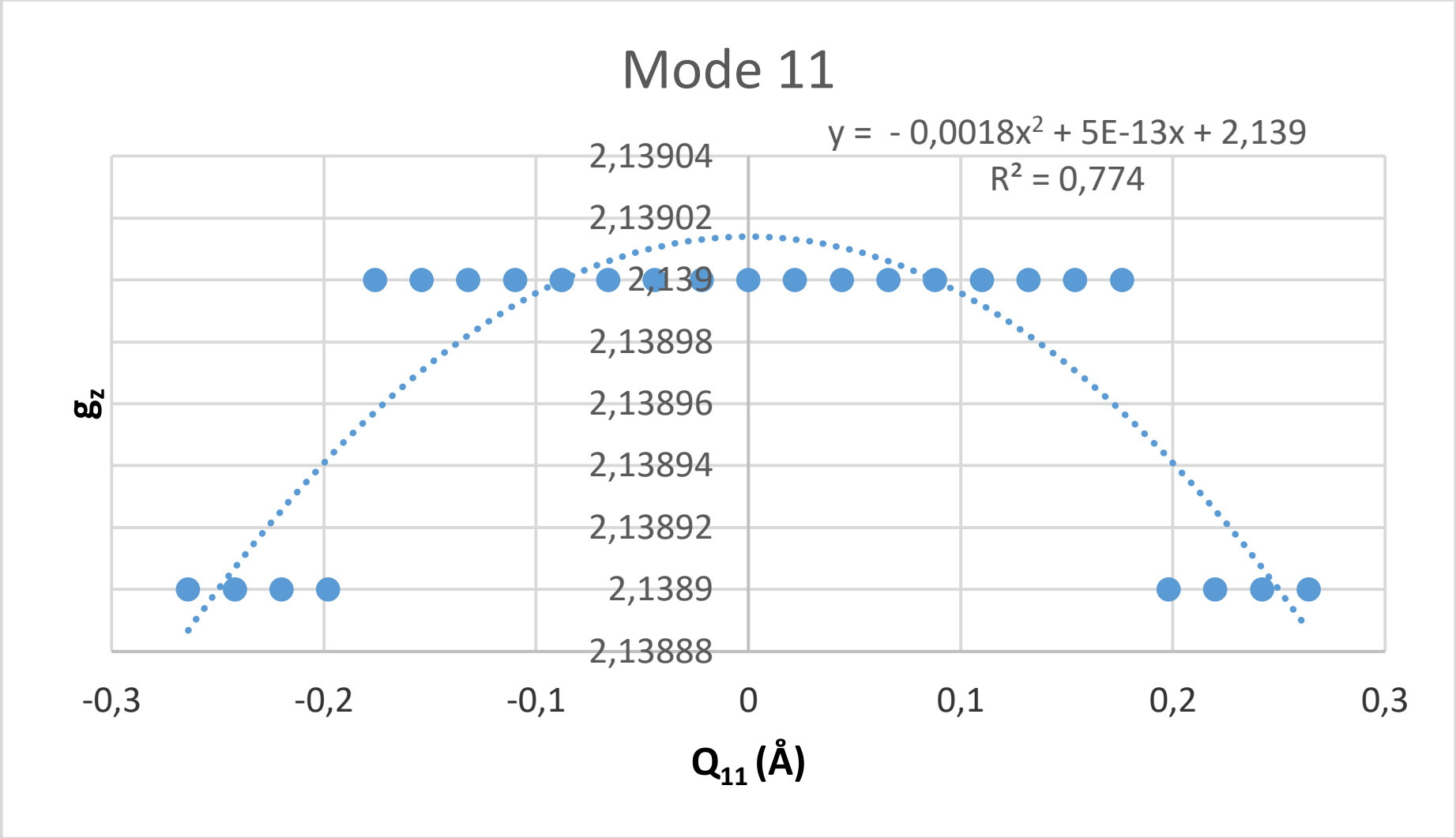
Mode 11
y = - 0,0018x² + 5E-13x + 2,139
R² = 0,774
2,13904
2,13902
2,139
2,13898
2,13896
2,13894
2,13892
2,1389
2,13888
gz
-0,3
-0,2
-0,1
0
0,1
0,2
0,3
Q11 (Å)

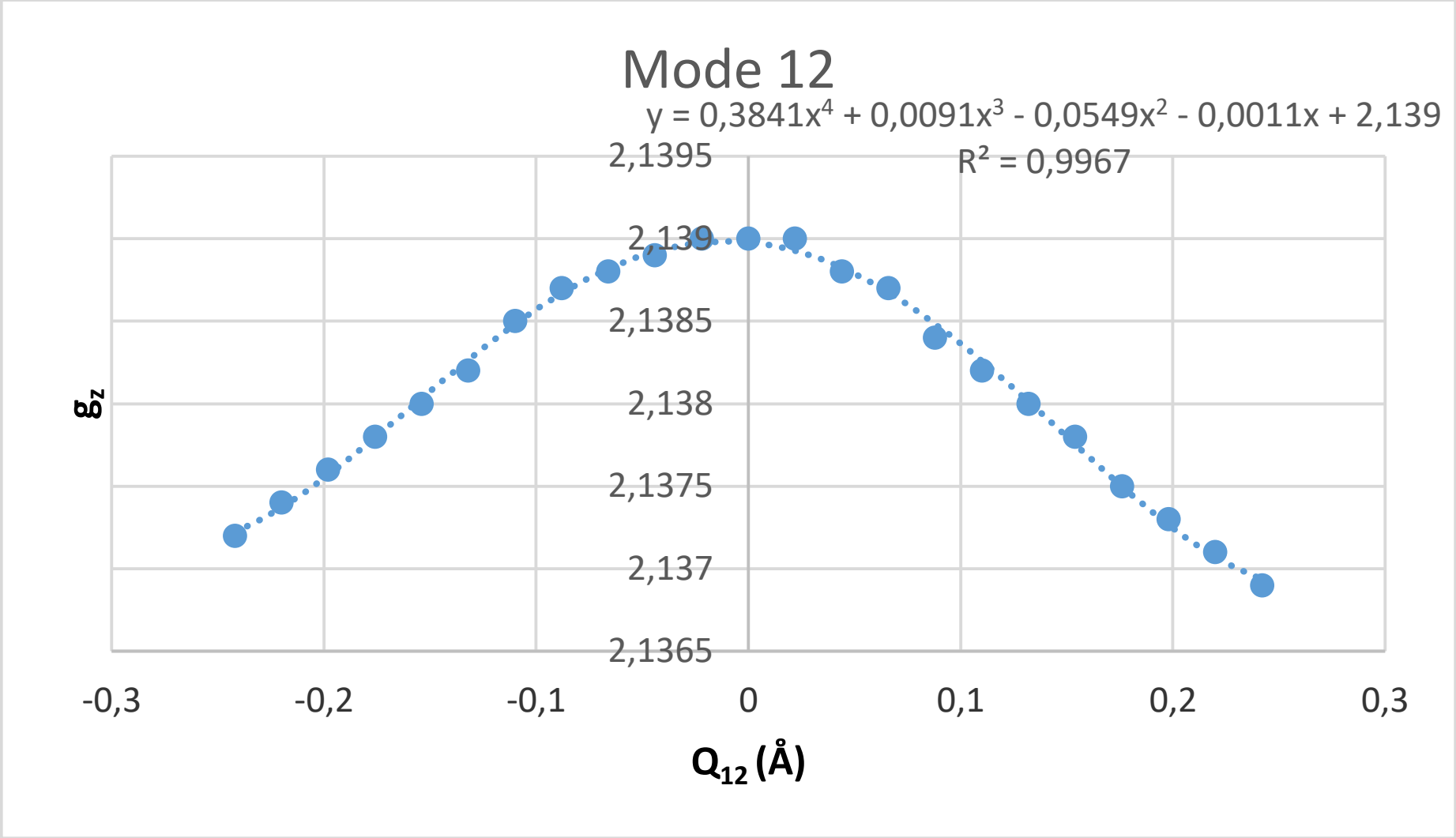
Mode 12
y = 0,3841x4 + 0,0091x3 - 0,0549x2 - 0,0011x + 2,139
R² = 0,9967
2,1395
2,139
2,1385
2,138
2,1375
2,137
2,1365
gz
-0,3
-0,2
-0,1
0
0,1
0,2
0,3
Q12 (Å)

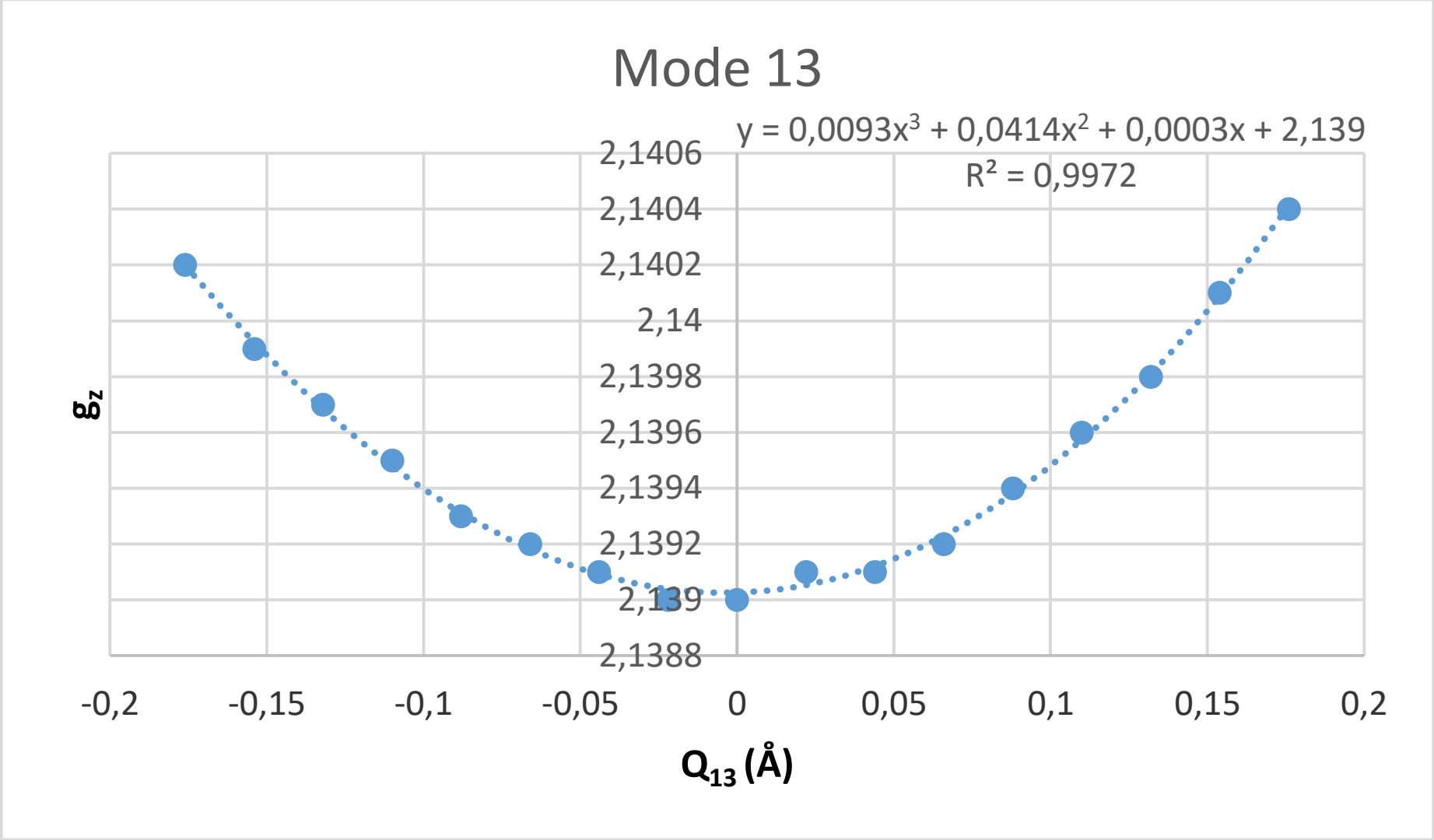
Mode 13
y = 0,0093x3 + 0,0414x2 + 0,0003x + 2,139
R² = 0,9972
2,1406
2,1404
2,1402
2,14
2,1398
2,1396
2,1394
2,1392
2,139
2,1388
gz
-0,2
-0,15
-0,1
-0,05
0
0,05
0,1
0,15
0,2
Q13 (Å)

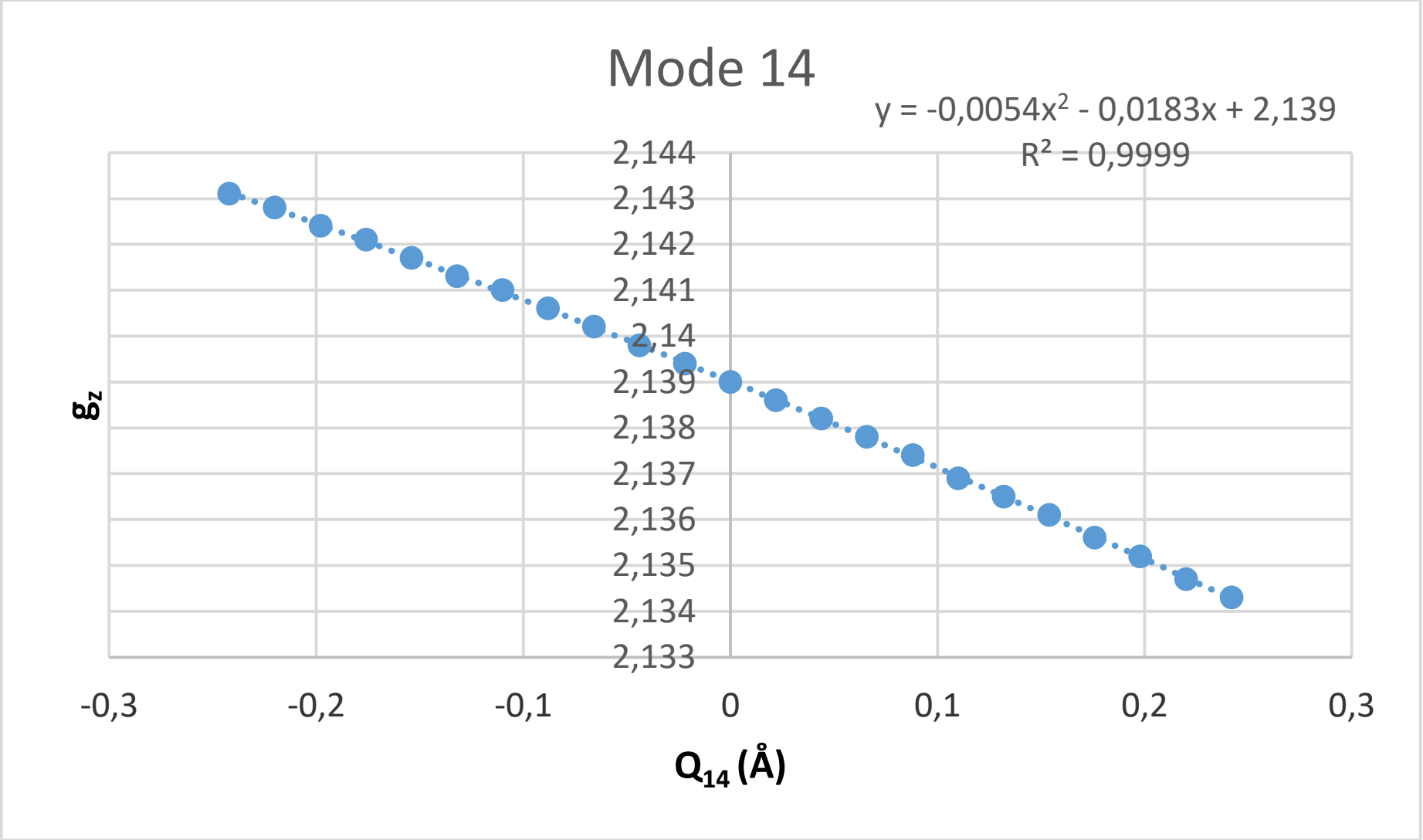

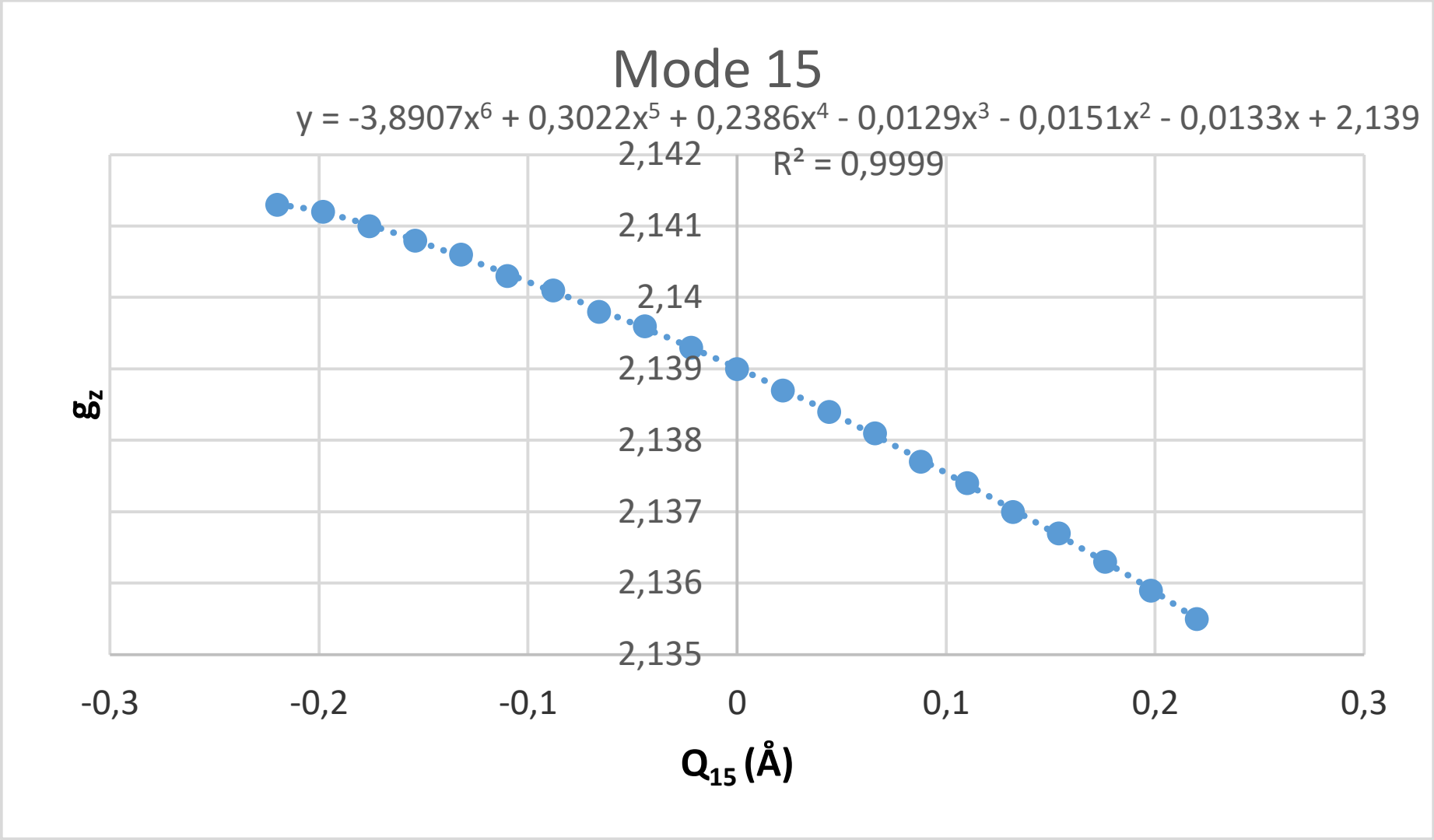

Mode 15
y = -3,8907x6 + 0,3022x5 + 0,2386x4 - 0,0129x3 - 0,0151x2 - 0,0133x + 2,139
R² = 0,9999
2,142
2,141
2,14
2,139
2,138
2,137
2,136
2,135
-0,3
-0,2
-0,1
0
0,1
0,2
0,3
gz
Q15 (Å)

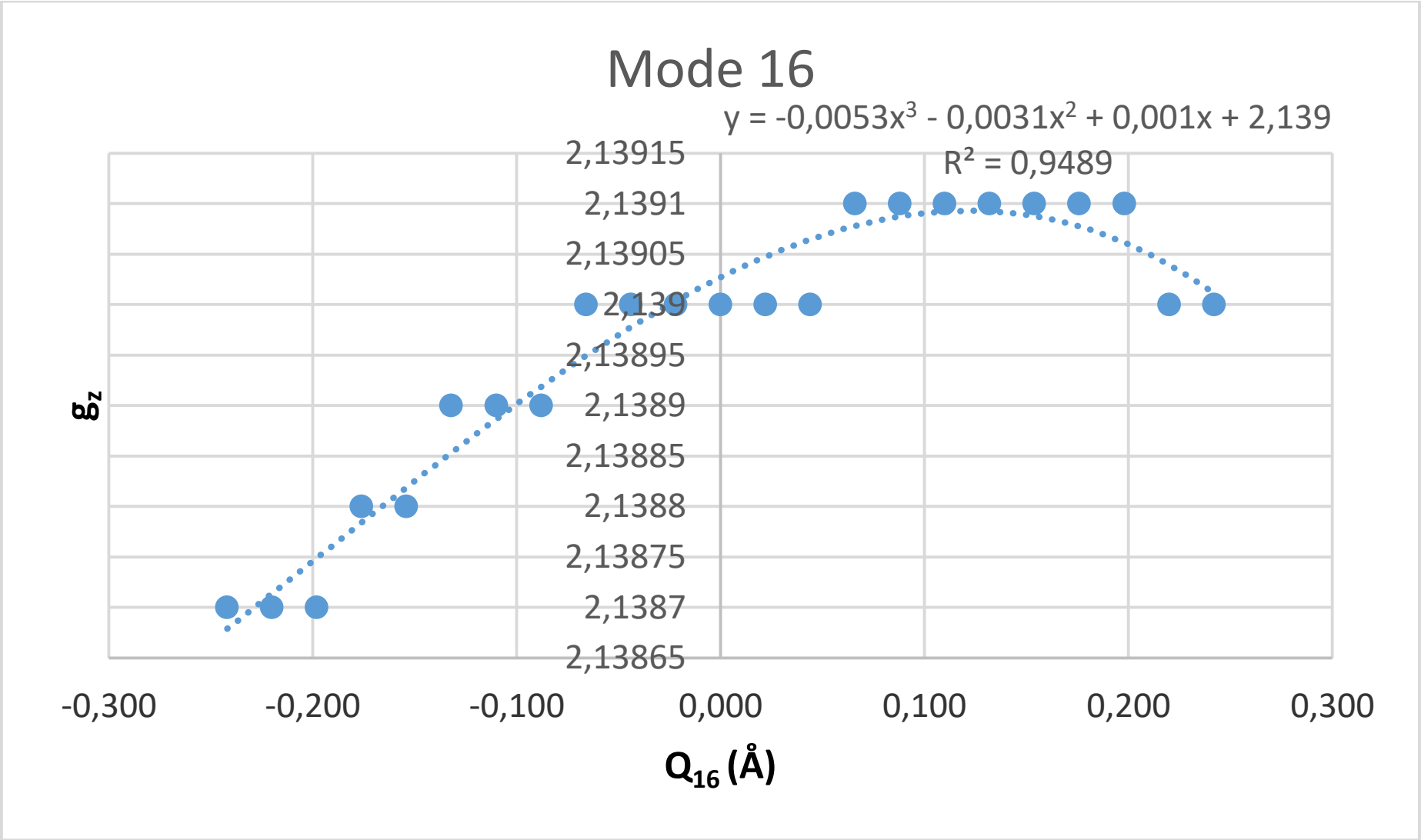

Mode 16
y = -0,0053x3 - 0,0031x2 + 0,001x + 2,139
R² = 0,9489
2,13915
2,1391
2,13905
2,139
2,13895
2,1389
2,13885
2,1388
2,13875
2,1387
2,13865
-0,300
-0,200
-0,100
0,000
0,100
0,200
0,300
gz
Q16 (Å)

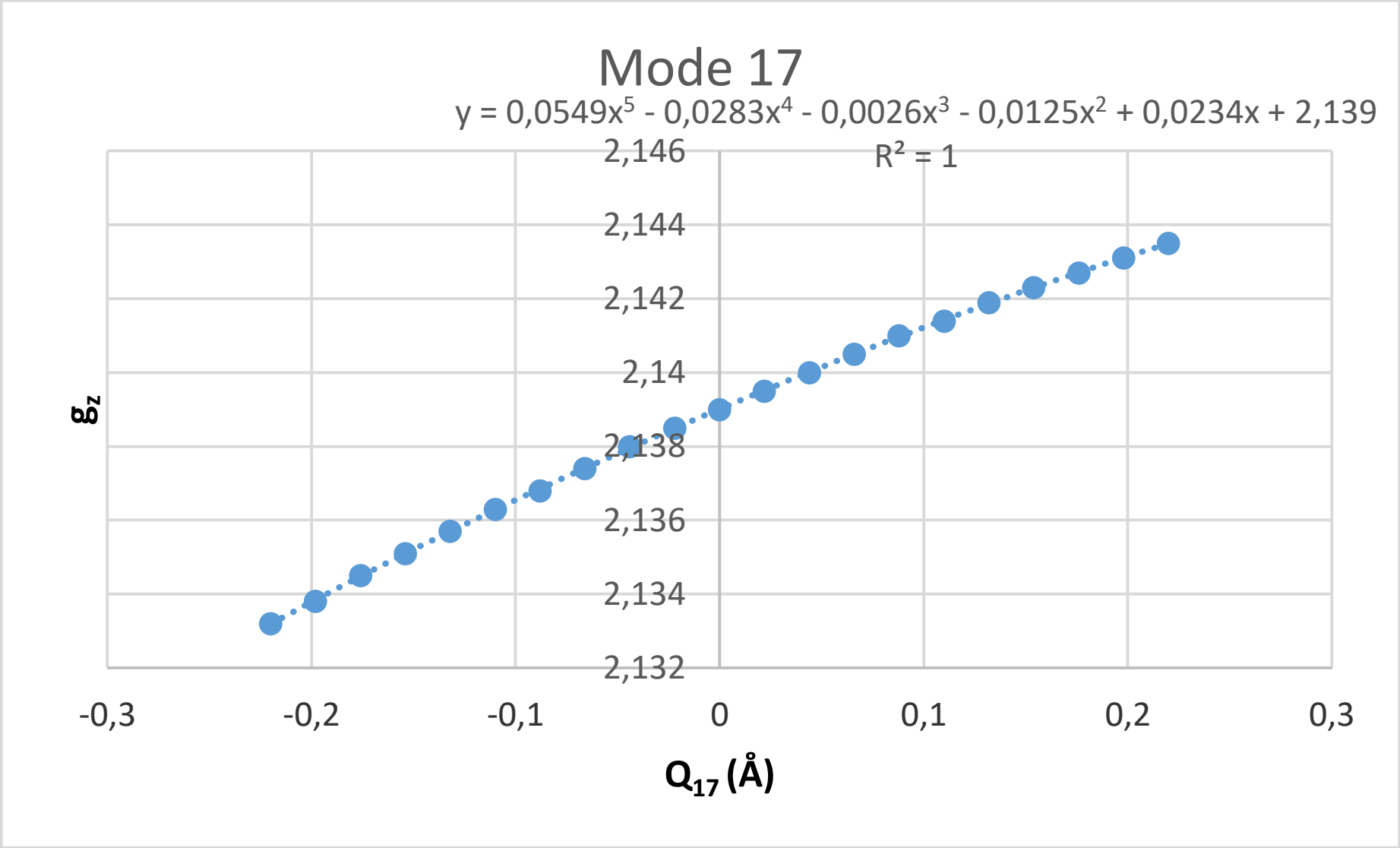

Mode 17
y = 0,0549x5 - 0,0283x4 - 0,0026x3 - 0,0125x2 + 0,0234x + 2,139
R² = 1
2,146
2,144
2,142
2,14
2,138
2,136
2,134
2,132
-0,3
-0,2
-0,1
0
0,1
0,2
0,3
gz
Q17 (Å)

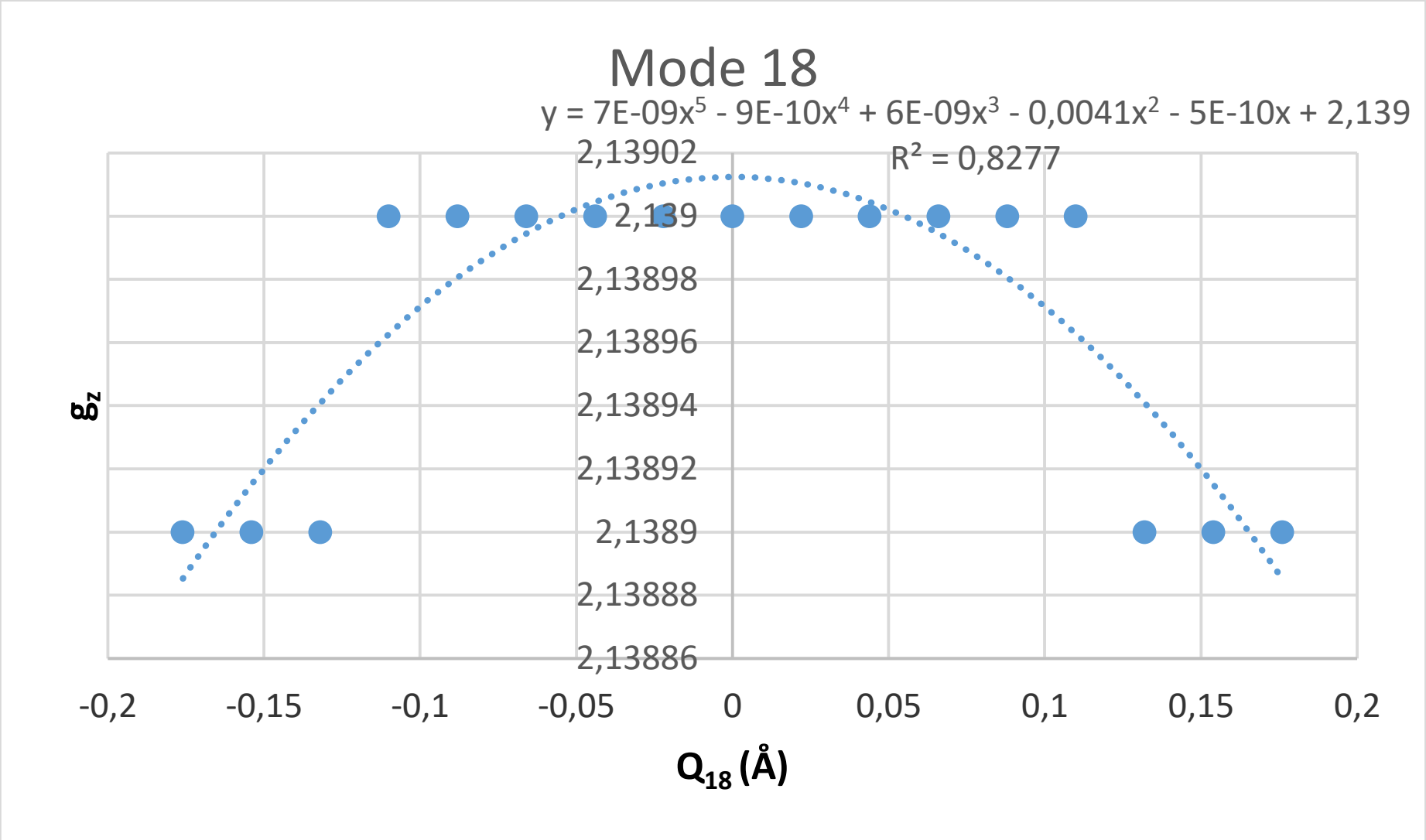
Mode 18
y = 7E-09x5 - 9E-10x4 + 6E-09x3 - 0,0041x2 - 5E-10x + 2,139
R² = 0,8277
2,13902
2,139
2,13898
2,13896
2,13894
2,13892
2,1389
2,13888
2,13886
gz
-0,2
-0,15
-0,1
-0,05
0
0,05
0,1
0,15
0,2
Q18 (Å)

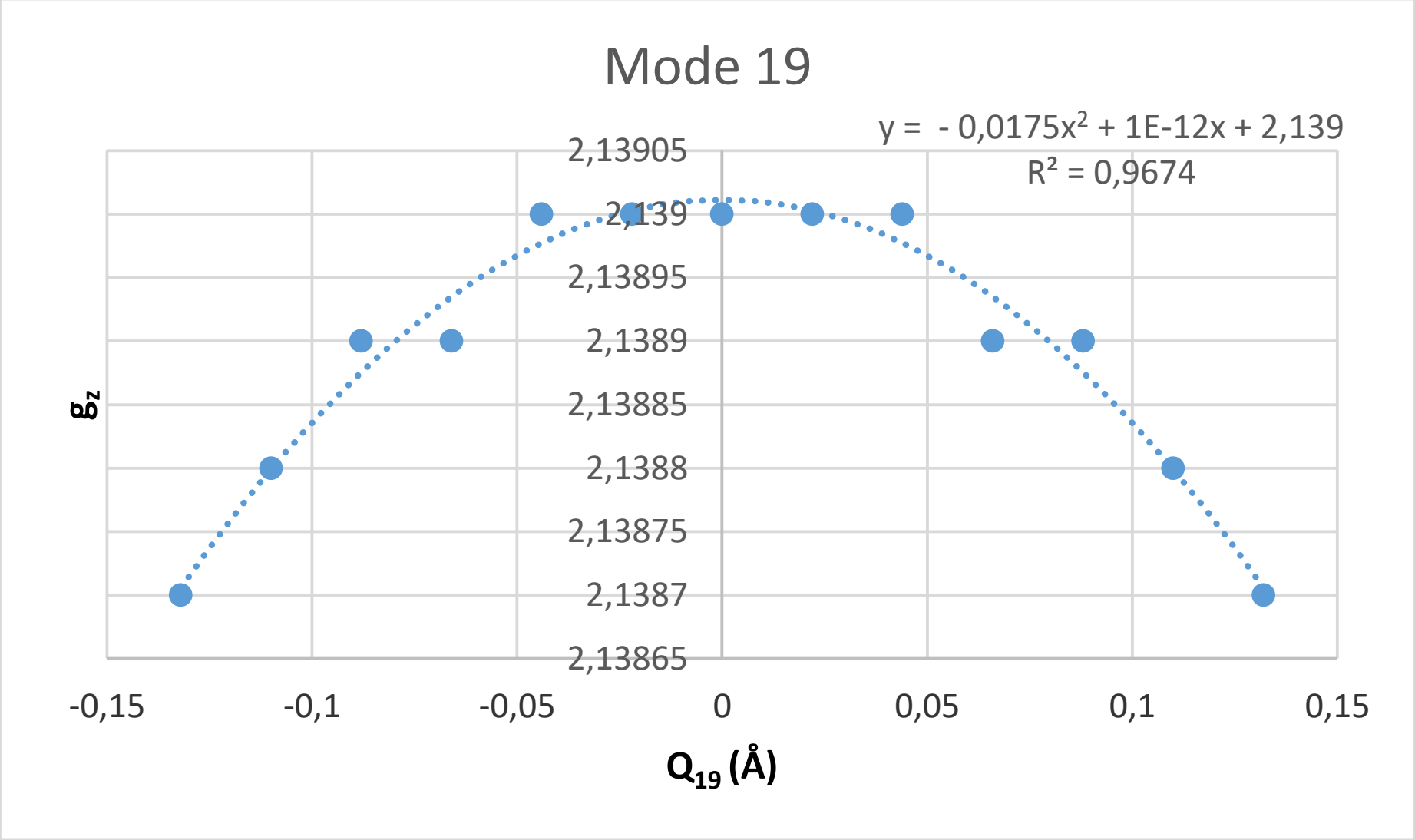
Mode 19
y = - 0,0175x2 + 1E-12x + 2,139
R² = 0,9674
2,13905
2,139
2,13895
2,1389
2,13885
2,1388
2,13875
2,1387
2,13865
gz
-0,15
-0,1
-0,05
0
0,05
0,1
0,15
Q19 (Å)

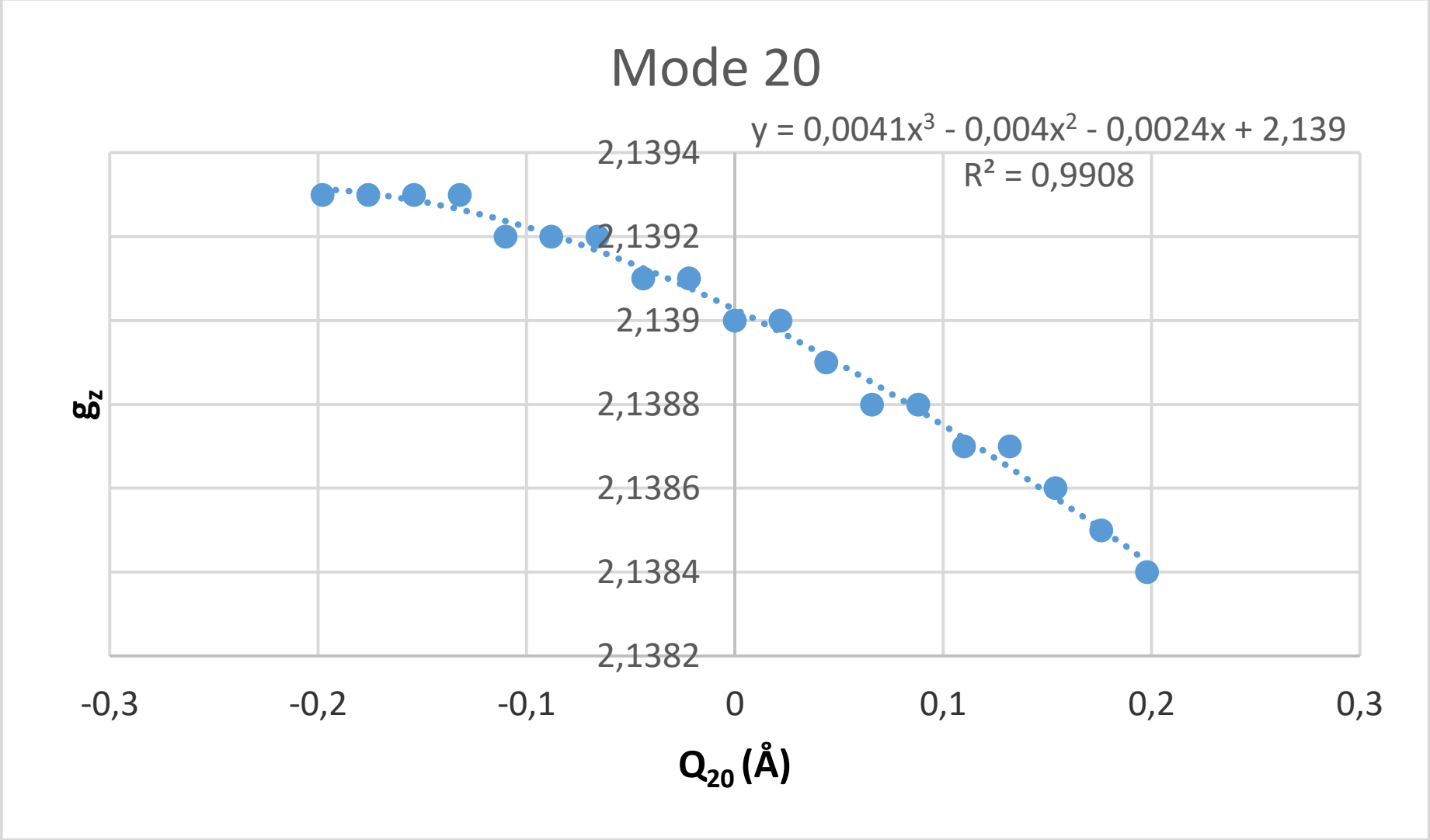
Mode 20
y = 0,0041x3 - 0,004x2 - 0,0024x + 2,139
R² = 0,9908
2,1394
2,1392
2,139
2,1388
2,1386
2,1384
2,1382
gz
-0,3
-0,2
-0,1
0
0,1
0,2
0,3
Q20 (Å)

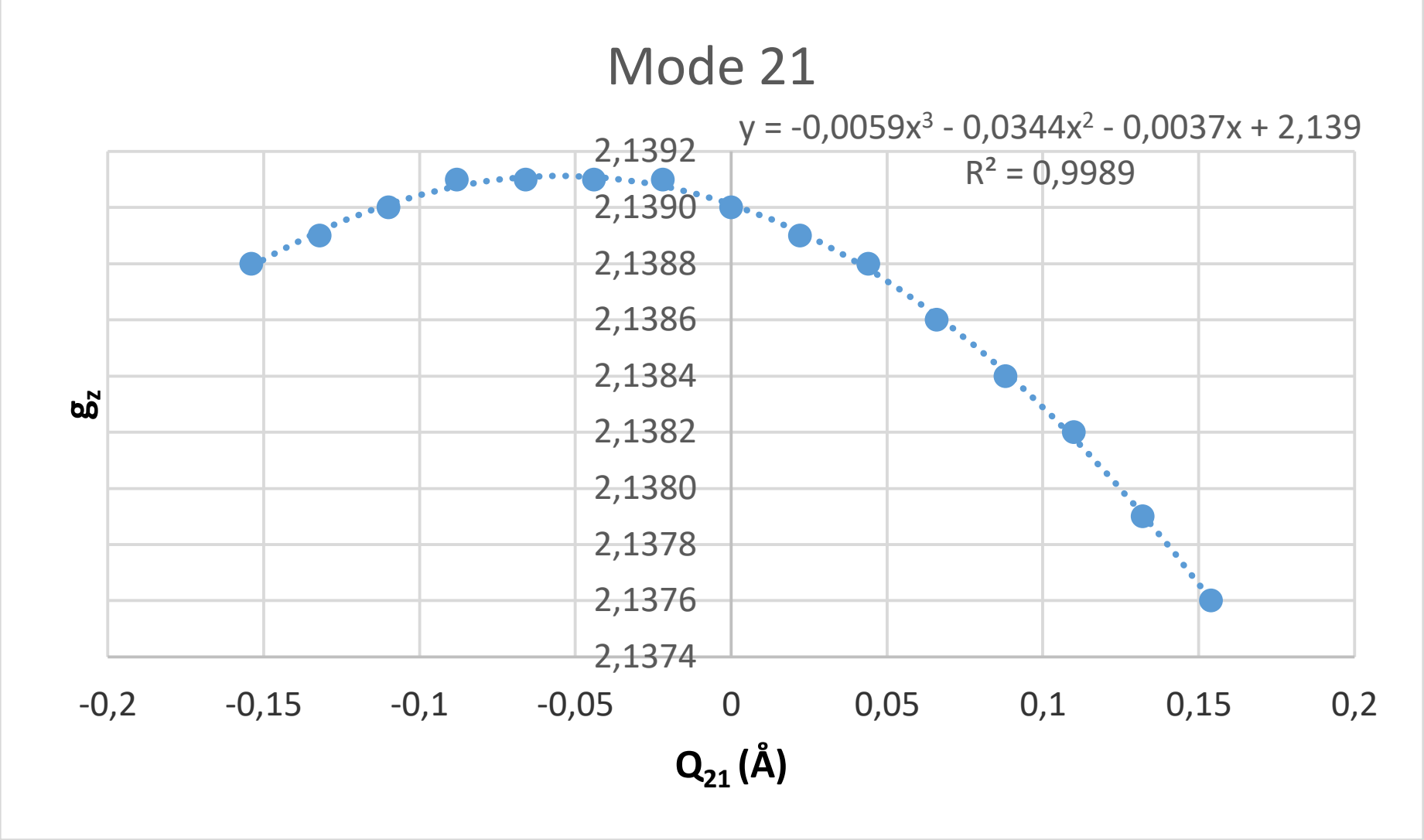
Mode 21
y = -0,0059x³ - 0,0344x² - 0,0037x + 2,139
R² = 0,9989
gz
Q21 (Å)

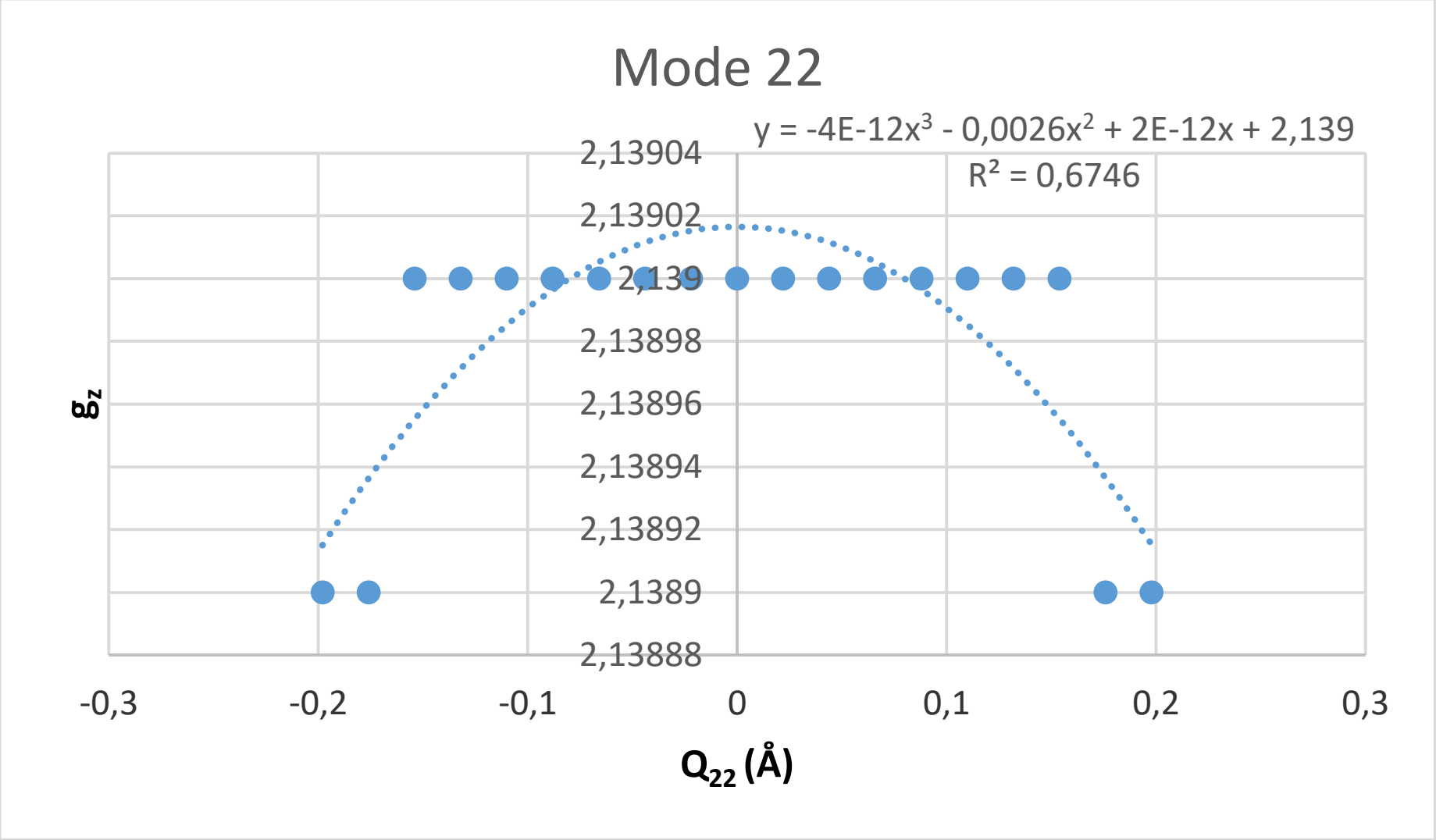
Mode 22
y = -4E-12x³ - 0,0026x² + 2E-12x + 2,139
R² = 0,6746
gz
Q22 (Å)

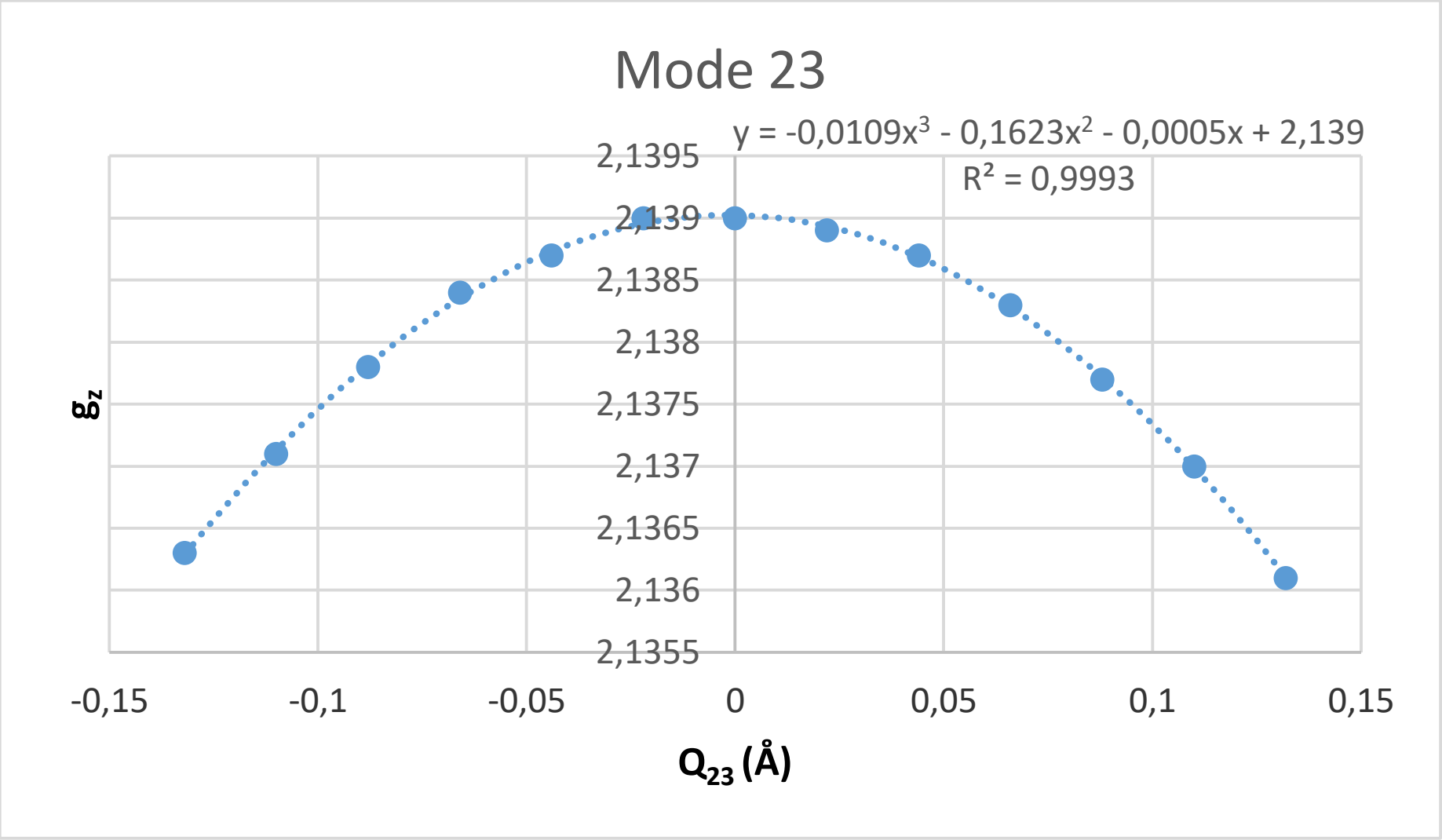
Mode 23
y = -0,0109x³ - 0,1623x² - 0,0005x + 2,139
R² = 0,9993
gz
Q23 (Å)

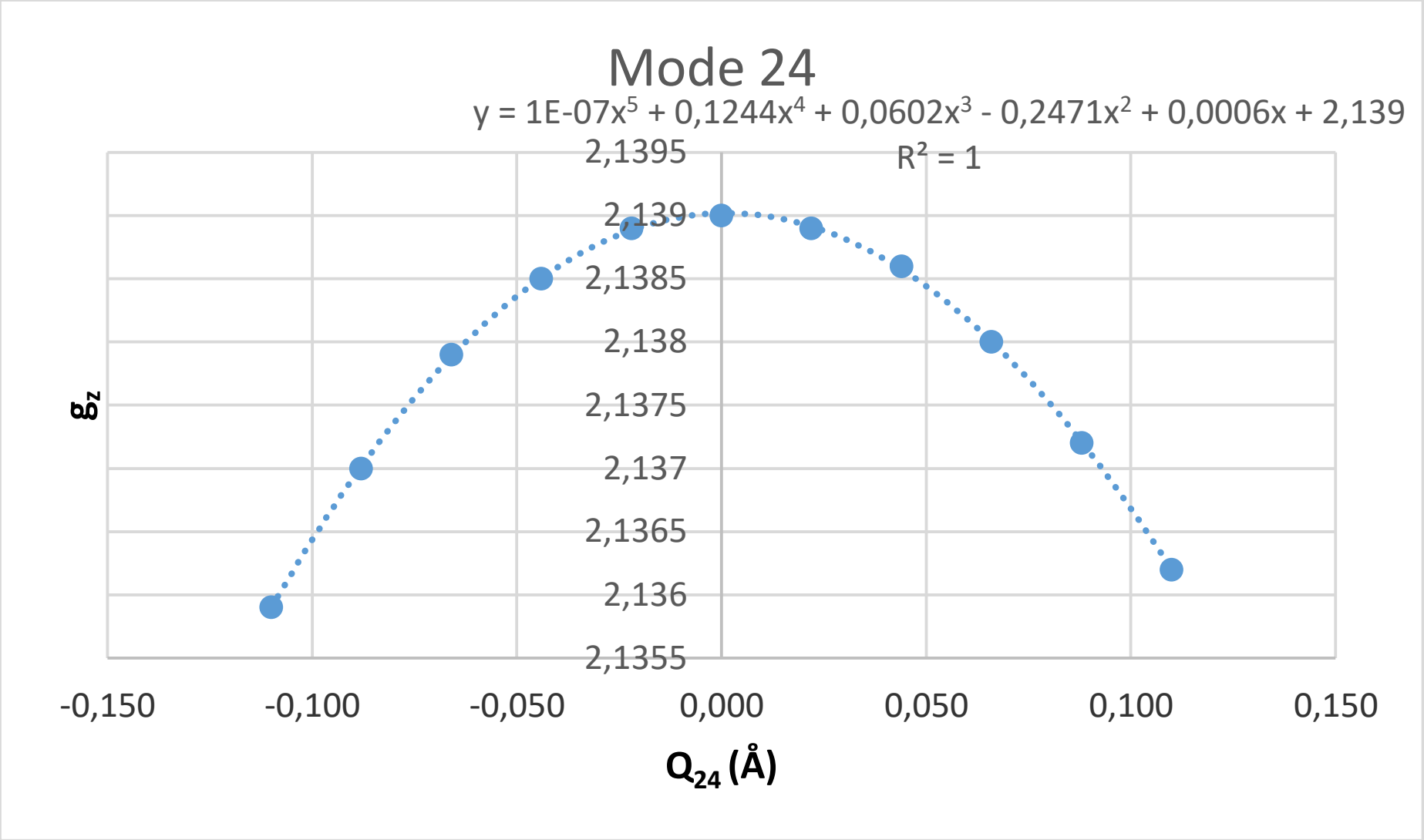

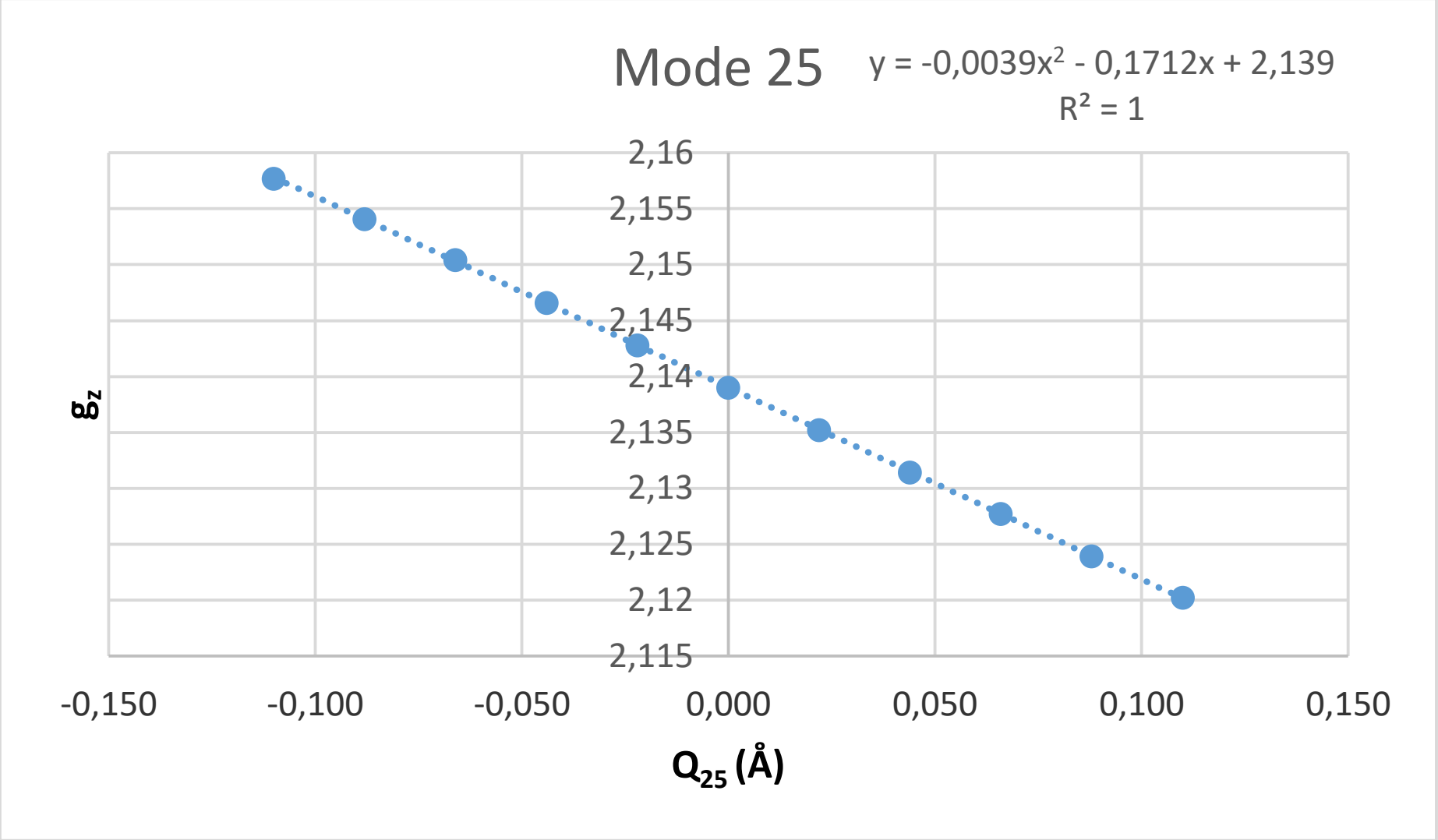

**7.** Table S3. List of the second derivatives "$(g_z)_{kk}$" of $g_z$ respect to the vibrational coordinate $Q_k$ for the first 25 vibrational modes of **1**.

| **Mode** | **1** | **2** | **3** | **4** | **5** |
|---|---|---|---|---|---|
| $(g_z)_{kk}$ ($Å^{-2}$) | 0.0024000 | 0.0020000 | 0.0028000 | 0.0190000 | 0.0050000 |
| **Mode** | **6** | **7** | **8** | **9** | **10** |
| $(g_z)_{kk}$ ($Å^{-2}$) | 0.0048000 | 0.0022000 | 0.0026000 | 0.0008000 | 0.0156000 |
| **Mode** | **11** | **12** | **13** | **14** | **15** |
| $(g_z)_{kk}$ ($Å^{-2}$) | -0.0036000 | -0.1098000 | 0.0828000 | -0.0108000 | -0.0302000 |
| **Mode** | **16** | **17** | **18** | **19** | **20** |
| $(g_z)_{kk}$ ($Å^{-2}$) | -0.0062000 | -0.0250000 | -0.0082000 | -0.0350000 | -0.0080000 |
| **Mode** | **21** | **22** | **23** | **24** | **25** |
| $(g_z)_{kk}$ ($Å^{-2}$) | -0.0688000 | -0.0052000 | -0.3246000 | -0.4942000 | -0.0078000 |

**8.** Figure S21. Pictures of some additional modes of **1** including vector displacements (blue arrows): 16 (first row) and 21 (second row). The pictures have been taken for some values of the distortion coordinate $Q_k$: -1.0 Å (left), +1.0 Å (right). Ligands with mobile parts should be avoided as they can help to distort the metal environment, an effect that can promote decoherence.

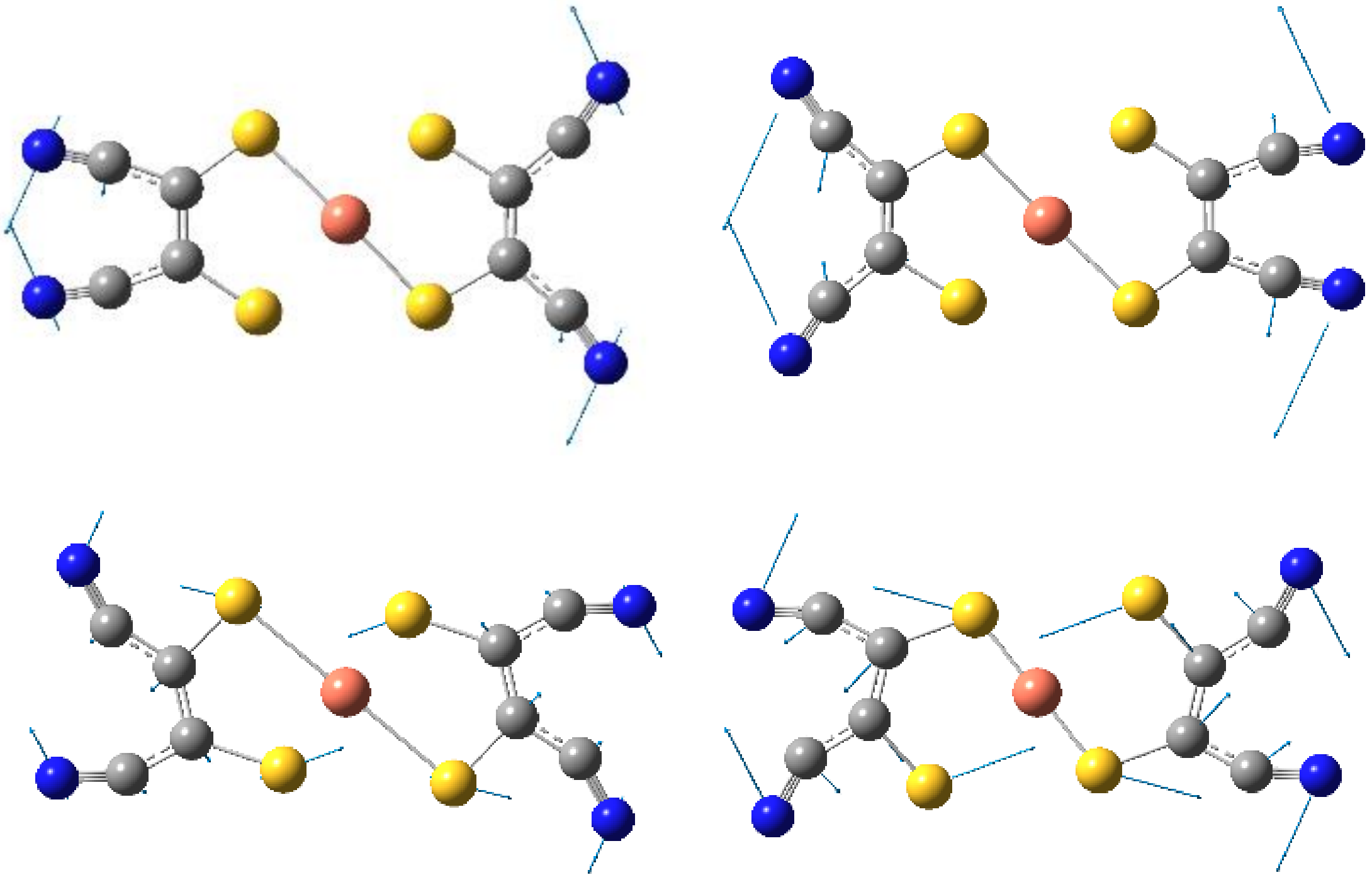

**9.** Equation S1. More practical $g_z$ statistical-averaged thermal expression, as given in the Eq. (4), (5) and (6) of the main paper, whose relevant parameters are now given in those units employed in a standard Gaussian09 package output. $N_A$ is the Avogadro's Number, and u.m.a.q. are chemical atomic mass units, the twelfth part of the mass of an atom of the isotope $^{12}C$. The distortion coordinates $Q_k$ are in angstroms.

$$\overline{\langle g_z \rangle}(T) \approx$$

$$\approx \overline{\langle g_z \rangle}(T=0) + \sum_{k=1}^{3P-6}\left[10^{20}\frac{5}{2\pi}\frac{N_A\cdot\hbar(J\cdot s)}{c(m/s)}\left(\frac{\partial^2 g_z}{\partial Q_k^2}\right)_e\left(\dot{A}^{-2}\right)\frac{1}{m_k(u.m.a.q.)\nu_k\left(cm^{-1}\right)}\langle n_k\rangle\right]$$

$$\overline{\langle g_z \rangle}(T=0) =$$

$$\left(g_z\right)_e + 10^{20}\frac{5}{4\pi}\frac{N_A\cdot\hbar(J\cdot s)}{c(m/s)}\sum_{k=1}^{3P-6}\left[\left(\frac{\partial^2 g_z}{\partial Q_k^2}\right)_e\left(\dot{A}^{-2}\right)\frac{1}{m_k(u.m.a.q.)\nu_k\left(cm^{-1}\right)}\right]$$

$$\langle n_k \rangle = \frac{1}{e^{\frac{\nu_k\left(cm^{-1}\right)}{k_B\left(cm^{-1}/K\right)T(K)}}-1}$$

The above expression is originated by considering a Grand Canonical Ensemble, where the probability $P$ of each single-molecule expectation value $\langle g_z \rangle^N$, characterized by the set *N*={*n₁*,…,*n_R*} of harmonic vibrational quantum numbers is given as:

$$P = \frac{e^{-\frac{E_1^{n_1}+\ldots+E_{3N-6}^{n_{3N-6}}}{k_B T}}}{Z}$$

being $E_k^{n_k}$ the harmonic vibrational energy of the mode k and $Z$ the Grand Partition Function:

$$Z = \sum_{n_1=0}^{\infty} \cdots \sum_{n_{3N-6}=0}^{\infty} e^{-\frac{E_1^{n_1}+\ldots+E_{3N-6}^{n_{3N-6}}}{k_B T}}$$

Then, the thermal dependence of the expectation value of B is obtained as follows, replacing $\langle g_z \rangle^N$ by Eq.(3) in the main paper:

$$\overline{\langle g_z \rangle}(T) = \frac{\sum_{n_1=0}^{\infty} \cdots \sum_{n_{3N-6}=0}^{\infty} e^{-\frac{E_1^{n_1}+\ldots+E_{3N-6}^{n_{3N-6}}}{k_B T}} \langle g_z \rangle^N}{\sum_{n_1=0}^{\infty} \cdots \sum_{n_{3N-6}=0}^{\infty} e^{-\frac{E_1^{n_1}+\ldots+E_{3N-6}^{n_{3N-6}}}{k_B T}}}$$

**10.** Figure S22. Calculated Relative Thermal Evolution $\Delta g_z$ in the temperature range 5K – 300K of **1** from Eq. (9) in the main paper.

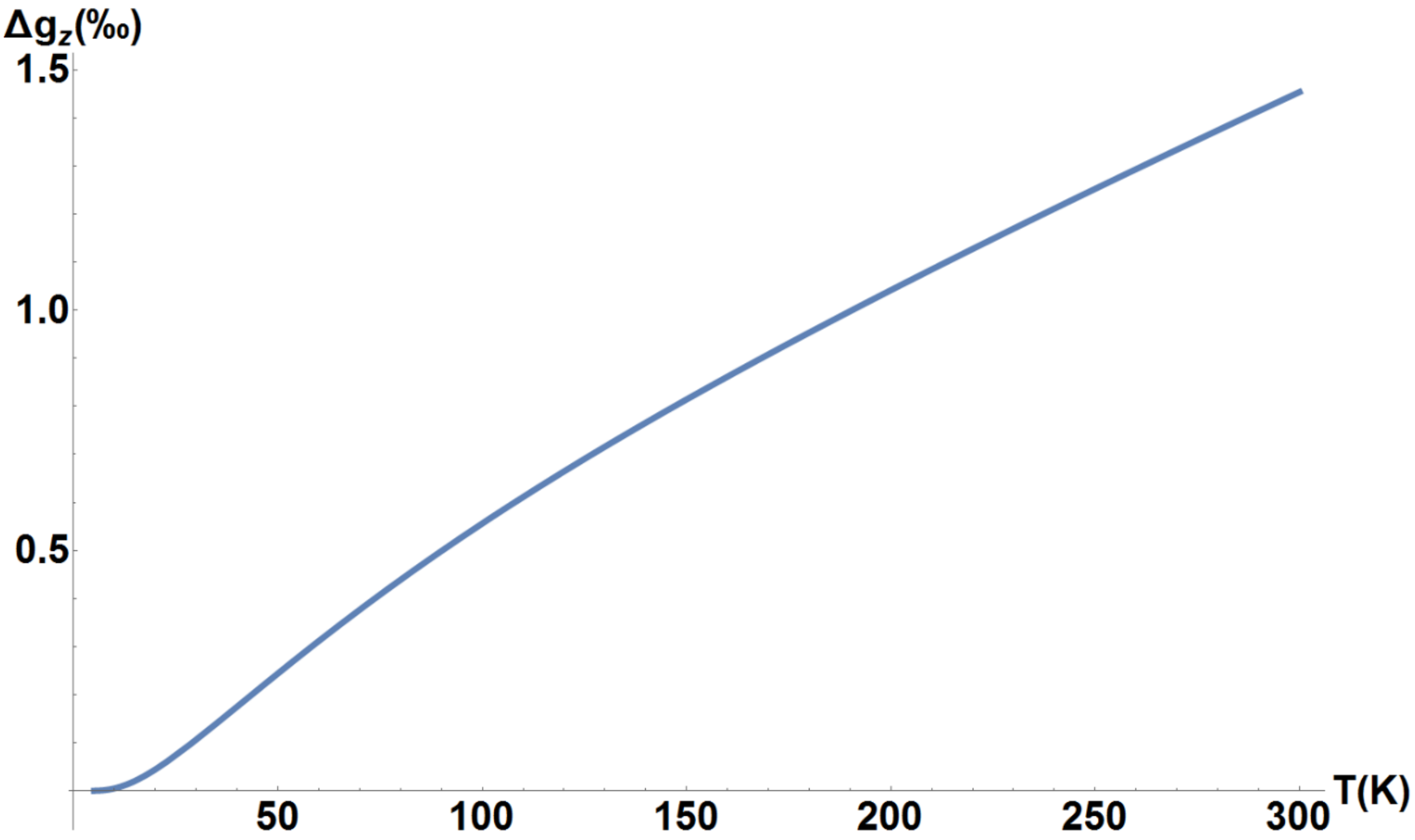

**11.** Boson number behavior at high temperatures.

The expression of the boson number is given by Eq. (6) in the main text. We can expand the exponential function by means of a McLaurin series: $e^x = 1 + x + x^2/2! + \dots + x^n/n! + \dots$ . In this case, $x = \nu_k/k_BT$. At high temperatures, x is small, and we can neglect second or higher order terms in the expansion: $e^x \approx 1 + x$. Then, the boson number becomes $1/(1 + x - 1) = 1/x = (k_B/\nu_k)T$, a straight line with slope $k_B/\nu_k$.

**12.** Supplementary Bibliography